\documentclass[twocolumn,prx,aps,eqsecnum]{revtex4-1}
\usepackage[T1]{fontenc}
\usepackage[latin9]{inputenc}
\usepackage{color}
\usepackage{bm}
\usepackage{amsmath}
\usepackage{amssymb}
\usepackage{graphicx}
\usepackage{braket}
\usepackage{esint}
\PassOptionsToPackage{normalem}{ulem}
\usepackage{ulem}
\usepackage[unicode=true,bookmarks=false,breaklinks=false,pdfborder={0 0 1},backref=false,colorlinks=true]{hyperref}
\hypersetup{pdfborderstyle=}

\makeatletter

\usepackage{color}
\usepackage{bm}
\usepackage{bbold}
\usepackage{soul}
\usepackage{epsfig}
\usepackage{verbatim}
\usepackage{array}
\newcommand{\be}{\begin{equation}}
\newcommand{\ee}{\end{equation}}
\newcommand{\bea}{\begin{eqnarray}}
\newcommand{\eea}{\end{eqnarray}}

\definecolor{kyrylo}{rgb}{0.7,0.5,0.2}
\definecolor{rein}{rgb}{0.2,0.5,0.8}
\definecolor{notneed}{rgb}{0.3,0.7,0.9}

\makeatother

\begin{document}
\global\long\def\sp#1#2{\langle#1|#2\rangle}%
\global\long\def\abs#1{\left|#1\right|}%
\global\long\def\avg#1{\langle#1\rangle}%
\global\long\def\ket#1{|#1\rangle}%
\global\long\def\bra#1{\langle#1|}%

\title{Towards dark space stabilization and manipulation in driven dissipative Majorana platforms  }
\author{Matthias Gau,$^{1,2}$ Reinhold Egger,$^{1}$ Alex Zazunov,$^{1}$ and Yuval Gefen$^{2}$}
\affiliation{$^{1}$~Institut f\"ur Theoretische Physik, Heinrich-Heine-Universit\"at,
D-40225 D\"usseldorf, Germany\\
$^{2}$~Department of Condensed Matter Physics, Weizmann Institute,
Rehovot, Israel 
}
\date{\today}
\begin{abstract}
We propose driven dissipative Majorana platforms for the stabilization and manipulation of robust quantum states. For Majorana box setups,  in the presence of environmental electromagnetic noise and with tunnel couplings to quantum dots, we show that
 the time evolution of the Majorana sector is governed by a Lindblad master equation over a wide parameter regime.
For the single-box case, arbitrary pure states (`dark states') can be stabilized by adjusting suitable gate voltages.
For devices with two tunnel-coupled boxes, we outline how to engineer dark spaces, i.e., manifolds of degenerate dark states, and how to stabilize fault-tolerant Bell states. 
The proposed Majorana-based dark space platforms rely on the constructive interplay of topological protection mechanisms and the autonomous quantum error correction
 capabilities of engineered driven dissipative systems.  Once a working Majorana platform becomes available, only standard hardware requirements  
are needed to implement our ideas.
\end{abstract}
\maketitle

\section{Introduction}\label{sec1}

It has been known for a long time that the dynamics of open quantum systems subject to external driving forces and coupled to environmental modes (`heat bath') can be described by master equations \cite{Weiss2007,Breuer2006,Gardiner2004}.
For a Markovian bath, the memory time of the bath represents the shortest time scale of the problem. The master equation is then of 
Lindblad type \cite{Lindblad1976,Lindblad1983}, where a Hamiltonian describes the coherent time evolution of the system's density matrix and a Lindbladian captures the dissipative dynamics.  (We here use `Lindbladian' for the dissipator terms in the master equations below.)  The Lindblad equation is the most general Markovian master equation which preserves the trace and positive semi-definiteness of the density matrix.

A major development over the past two decades has come from the realization that driven dissipative (DD) quantum systems can be stabilized in a pure quantum state by appropriate engineering of the driving fields and of the coupling to the dissipative environment \cite{Plenio1999,Beige2000,Plenio2002,Diehl2008,Kraus2008,Diehl2010,Diehl2011,Bardyn2013,Zanardi2014,Albert2014,Jacobs2014,Albert2016,Goldman2016,Wiseman2010}. 
Such states are eigenstates of the corresponding Lindbladian with zero eigenvalue, i.e., the operation of the Lindbladian leaves them inert. We therefore will refer to these DD stabilized states 
as \emph{dark states} in what follows. Rather than viewing the coupling to a dissipative environment as foe (e.g., leading to decoherence of quantum states and undermining the utilization of  similar platforms for quantum information processing), the combined effect of drive and dissipation can thus be harnessed to engineer quantum-coherent pure states. 
Going beyond dark states, the stabilization of a \emph{dark space} \cite{Iemini2015,Iemini2016,Santos2020} --- a manifold spanned by multiple degenerate dark states --- 
raises the prospects of employing such systems as viable platform for quantum information processing.  
Reference~\cite{Touzard2018} reports on recent experimental results in this direction.

Using trapped ions or superconducting qubits, the above ideas have already allowed for first qubit stabilization experiments  \cite{Geerlings2013,Lu2017,Touzard2018}, for
the implementation of quantum simulators  \cite{Barreiro2011,Schindler2013}, and for the generation of selected highly entangled multi-particle states \cite{Shankar2013,Leghtas2013,Reiter2016,Liu2016}.  Systems composed of many coupled qubits stabilized by DD mechanisms could eventually result in 
universal quantum computation platforms \cite{Verstraete2009,Fujii2014}, where fault tolerance is the consequence of
autonomous error correction \cite{Terhal2015} due to the engineered dissipative environment, without the need for active feedback \cite{Wiseman2010,Kerckhoff2010,Murch2012,Kapit2015,Kapit2016}.  
Recent experimental progress on autonomous error correction in DD qubit systems has been described in Refs.~\cite{Leghtas2013,Liu2016,Reiter2017,Puri2019}.
At present, reported fidelities in DD qubit setups (which by construction are stable in time) are typically below 90$\%$ for state stabilization, with significantly lower fidelities
for single- or two-qubit gate operations. 

Another important and at first glance unrelated development towards the (so far elusive) goal of fault-tolerant universal quantum computation comes from the field of topological quantum computation \cite{Nayak2008}.
By using topological quasiparticles \cite{Wen2017} for encoding and processing quantum information, the latter is nonlocally distributed in space
and thereby protected against local environmental fluctuations.  In general terms,
for practically useful and scalable DD systems with multiple degenerate dark states,  the coupling to the environment has to  be carefully engineered such that it is blind to all system operators acting within the targeted dark space manifold \cite{Facchi2000}. 
It will thus be imperative to avoid residual (uncontrolled and unwanted) noise sources. 
In that regard, platforms harboring topological quasiparticles may offer a key advantage since they should come with a strongly
reduced intrinsic sensitivity to residual environmental fluctuations as compared to conventional systems. 
The simplest candidate for topological quasiparticles is given by
Majorana bound states (MBSs), which are localized zero-energy states in topological superconductors. For Majorana reviews, see Refs.~\cite{Alicea2012,Leijnse2012,Beenakker2013,Sarma2015,Aguado2017,Lutchyn2018,Zhang2019a}.
Topological codes relying on MBSs have so far been discussed in the context of active error correction \cite{Alicea2011,Terhal2012,Hyart2013,Vijay2015,Aasen2016,Landau2016,Plugge2016,Plugge2017,Karzig2017,Litinski2017,Wille2019},
where periodically repeated stabilizer measurements are needed for fault tolerance.
It remains an important challenge to devise feasible and scalable Majorana platforms exploiting passive error correction strategies, 
where DD mechanisms serve to continuously measure the system in a way that the desired highly entangled many-body quantum state becomes stabilized automatically, see, e.g., Ref.~\cite{Herold2017}.  
While this ambitious goal is beyond the scope of our work, we here analyze related questions for DD systems with up to eight MBSs.  

For a mesoscopic floating (not grounded) topological superconductor harboring four MBSs, strong  charging effects 
\cite{Fu2010} imply that the ground state is doubly degenerate under Coulomb valley conditions (see Sec.~\ref{sec2a} for details).  
Such a superconducting island is therefore a good candidate for a topologically protected Majorana qubit,
named Majorana box qubit \cite{Plugge2017} or tetron \cite{Karzig2017}.  Thanks to the nonlocal Majorana encoding of quantum information, such a qubit allows 
for unique addressability options via electron cotunneling when quantum dots (QDs) or normal leads are attached to the island by tunneling contacts, see also Refs.~\cite{Gau2018,Munk2019}. 
 Majorana qubits have not yet been experimentally realized. However, the recent emergence of new Majorana platforms (see, e.g., 
 Refs.~\cite{Liu2018b,Zhang2018b,Wang2018b,Sajadi2018,Ghatak2018,Murani2019}) in addition to the semiconductor nanowire platform mainly explored 
so far \cite{Lutchyn2018,Zhang2019a} indicates that they may be available in the foreseeable future.   We note that alternative Majorana qubit designs have been put forward, e.g., in Refs.~\cite{Terhal2012,Hyart2013,Aasen2016}. Many of the ideas discussed below can be adapted to those setups as well. 

\subsection{Motivation and goals of this work}

We here show that once available, Majorana box devices yield highly attractive platforms for implementing DD protocols
aimed at the realization of dark states and/or dark spaces.  The driving field is applied to the tunnel link connecting a pair of QDs, and 
dissipation is due to environmental electromagnetic noise.  
To the best of our knowledge, apart from a distantly related proposal for the
DD stabilization of Majorana-based quantum memories  \cite{Bardyn2016},
no studies of DD Majorana systems have appeared in the literature so far.
We note that the DD engineering of MBSs in cold-atom based Kitaev chains \cite{Diehl2011,Bardyn2013,Goldman2016} differs from our ideas:  We consider topological superconductors harboring native MBSs, and then subject the resulting Majorana systems to DD stabilization
and manipulation protocols targeting dark states and/or dark spaces. Our unique platform enables us to employ QDs as external knobs to be used not only for state engineering but also for state manipulation. 

Our motivation for designing and studying novel DD stabilization and manipulation schemes using Majorana platforms rests on several arguments and expectations:
\begin{enumerate}
\item
 Since uncontrolled environmental effects are largely  suppressed by topological protection mechanisms, one may  reach higher
fidelities than those reported so far for DD dark state or dark space implementions using conventional (topologically trivial) platforms.  This point should be especially important for high-dimensional dark spaces, where residual noise effects could
break the degeneracy of the dark states spanning the dark space manifold \cite{Facchi2000}.  Such spaces are highly attractive candidates for implementing fault-tolerant quantum computing platforms. These topological protection elements are especially important for platforms where the Lindblad spectrum is not gapped.  

\item It is known that for large-scale Majorana surface codes, where active feedback is needed for code stabilization,  
the fault-tolerance error threshold is much more benign than for conventional bosonic surface codes, see Refs.~\cite{Vijay2015,Plugge2016,Fowler2012} and references therein.
In particular, in Majorana surface codes no ancilla qubits are needed for stabilizer readout at all.
We expect that our dark space constructions using MBS systems can allow for similar fault tolerance advantages over conventional dark space realizations.   
However, more work is needed to reach a quantitative conclusion on this point. 
\item The DD stabilization and manipulation of Majorana-based dark states or dark spaces offers several practical 
advantages. In particular, the robustness of such states as quantified by the dissipative gap is expected to be superior to 
quantum states that are encoded without DD mechanisms in native Majorana devices, see Sec.~\ref{sec3}.
Moreover, a small overlap between MBSs is often tolerable, without causing dephasing of  dark states, cf.~Sec.~\ref{sec3e}.

\item When steering a state into the dark space or manipulating a state within the dark space, one may need to maximize its purity, having in mind quantum information manipulation protocols. For this purpose, we may adiabatically switch on a suitable perturbation either to the Lindbladian dissipator or to the accompanying Hamiltonian, thereby breaking the degeneracy of 
the dark space.  In this manner, one can revert to a specific pure dark state, manipulate this state, and subsequently adiabatically switch off this perturbation again. 
The DD Majorana platforms discussed below offer convenient tools to switch on and off such degeneracy-breaking perturbations.
\end{enumerate}

\begin{figure}[t]
\begin{centering}
\includegraphics[width=0.9\columnwidth]{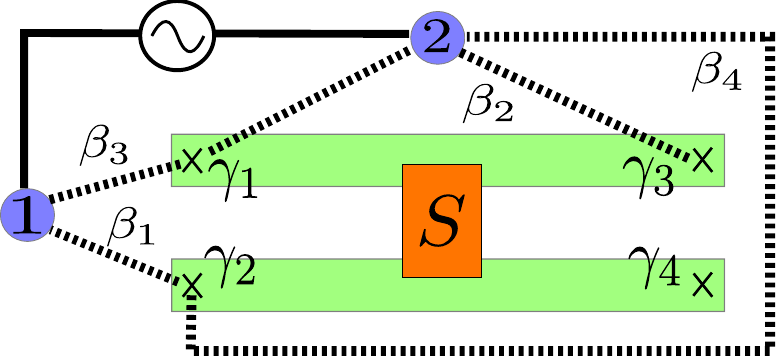}
\end{centering}
\caption{Schematic sketch of a driven dissipative Majorana box setup. The superconducting island harbors four Majorana operators $\gamma_\nu$, three of which are tunnel-coupled to two single-level quantum dots (QDs, in blue). 
The Majoranas could be realized as end states of two parallel topological superconductor nanowires (green) which are electrically connected by a superconducting bridge (orange) \cite{Plugge2017}.
The tunnel links connecting QDs to MBSs are shown as dashed lines. The phases $\beta_j$ in Eq.~\eqref{betadef} are also indicated. Due to the large box charging energy,
transport between different QDs through the Majorana island proceeds only via cotunneling processes.  
These cotunneling processes can be inelastic, involving the emission or absorption of photons from the dissipative electromagnetic environment.
In addition, a driving field can pump electrons via a tunnel link between the QDs (solid line).
 }
\label{fig1}
\end{figure}

The dynamics of the Majorana degrees of freedom in a device such as the one depicted in Fig.~\ref{fig1} will here be discussed on several conceptual levels. We show that our DD protocols indeed give rise to master equations of Lindblad type. These equations contain both a Hamiltonian (governing the unitary part of the time evolution) and a Lindbladian (causing dissipative dynamics).  By choosing suitable parameter values as discussed in Sec.~\ref{sec3},  we demonstrate that an arbitrary dark state can be stabilized. In more complex two-box devices, see Sec.~\ref{sec4}, the Lindbladian can be engineered to support a multi-dimensional dark space.
As a generic initial state is driven towards the dark space, we show (see also Ref.~\cite{ourprl}) how to optimize the purity, the fidelity (i.e., the overlap of the state with the target dark space), and the speed of approach.

A major benefit of applying  DD strategies to a topologically nontrivial system comes from the insight that  
one can here implement unidirectional cotunneling processes in an elementary and practically useful manner.  Using Majorana boxes which are tunnel-coupled to two quantum dots,
we show that the combination of driving fields, energy relaxation, and the tunability of tunneling amplitudes allows for the controlled design of 
directed cotunneling processes.  The latter directly determine the important jump operators in the Lindblad equation. 

In our accompanying short paper \cite{ourprl}, we provide a summary of our key ideas and apply them to  show that in a two-box setup, one can stabilize and manipulate 
`dark qubit' states.  In effect, the topologically protected native Majorana qubit discussed in Refs.~\cite{Plugge2017,Karzig2017} (which exists in a single box) is thereby stabilized by adding another protection layer due to DD mechanisms (at the prize of adding a second box).

\subsection{Overview}

In order to guide the focused reader through this long article, we here provide a short overview summarizing the content of the subsequent sections.
In addition,  Table~\ref{table1} summarizes the key symbols and notations used throughout this paper.
\begin{itemize}
\item In Sec.~\ref{sec2}, we introduce the theoretical concepts and physical ingredients needed for the
DD stabilization and manipulation of dark states using a single Majorana box, see Fig.~\ref{fig1}, and we derive the dynamical equations.  Our model is introduced in Sec.~\ref{sec2a},
 where the dissipation arises from environmental electromagnetic fluctuations and the drive is applied to a pair of QDs.
We subsequently derive the Lindblad equation \cite{Weiss2007,Breuer2006,Gardiner2004,Lindblad1976,Lindblad1983}  governing the time evolution of the combined QD-Majorana system in Sec.~\ref{sec2b}, 
where we also present numerical results for the dynamics obtained from this Lindblad master equation.
Remarkably, up to initial transient behaviors, one can describe the dynamics in the Majorana sector in terms of a reduced 
Lindblad equation, where the QD degrees of freedom have been traced out.  We describe this step in 
Sec.~\ref{sec2c}, along with a discussion of the conditions under which this reduced Lindblad equation applies.
All of our subsequent results are obtained by employing this reduced Lindblad equation.
\item In Sec.~\ref{sec3} we then describe dark state stabilization protocols for the single-box device in Fig.~\ref{fig1}.
We begin in Sec.~\ref{sec3a} with the case of Pauli operator eigenstates, followed by the stabilization of the so-called magic state in Sec.~\ref{sec3b}.
In Sec.~\ref{sec3c}, the role of increasing temperature on our stabilization protocols is examined. 
Interestingly, as shown in Sec.~\ref{sec3d}, we find that for certain parameter settings, dark states can be stabilized even in the absence of any drive.
However, the field-free stabilization is limited to very special conditions and is also characterized by rather small dissipative gaps. 
In practical implementations, it will thus be preferable to employ a driving field.
Finally, in Sec.~\ref{sec3e}, we discuss additional points, e.g., concerning the role of Majorana state overlaps or how to perform a parity readout of the stabilized states. 
\item In Sec.~\ref{sec4}, we  turn to a setup with two coupled boxes and present our DD stabilization and manipulation protocols for quantum states that belong to a dark space manifold.  The Lindblad equation for this setting is derived in Sec.~\ref{sec4a}.  We explain how one can engineer a degenerate dark space in Sec.~\ref{sec4b}. 
This topic is the main focus of Ref.~\cite{ourprl}, and the discussion is therefore kept rather short here.   Finally, in Sec.~\ref{sec4c}, 
we show how to stabilize Bell states in the two-box setting.
\item 
The paper concludes with a summary and an outlook in Sec.~\ref{secConc}.  
\end{itemize}
Technical details and additional information can be found in three Appendices. Let us also remark that we often use units with $\hbar=k_B=1$.

\begin{table*}[t]
\renewcommand{\arraystretch}{1.2}
\begin{center}
\begin{tabular}{c|l|c} \hline\hline
Symbol & Meaning & First appearance\\ \hline \hline 
\emph{Model parameters:} &  & \\ \hline
    $A$ & drive amplitude & \eqref{Hdriv}\\ \hline
    $\alpha$ & dimensionless system-bath coupling for Ohmic bath & \eqref{alphadef}\\ \hline
    $\beta_j$ & phases of the tunnel couplings $\lambda_{j\nu}$ & \eqref{betadef} \\ \hline
    $E_C$ & charging energy of the Majorana box & \eqref{Hbox} \\ \hline
     $\epsilon_{j}$ & level energy of the respective quantum dot & \eqref{HDots}\\ \hline
      $g_0$ & cotunneling scale for single-box setup, $g_0=t_0^2/E_C$ & \eqref{Heff}\\ \hline
       $\tilde g_0^{}$ & cotunneling scale for double-box setup  & \eqref{tildeg0}\\ \hline
    $\lambda_{j\nu}$ & tunnel coupling between QD fermion $d_j$ and Majorana operator $\gamma_{\nu}$ &   \eqref{hatlambda}\\ 
                     & (`state design parameters') & \\ \hline
   $M$ & number of MBSs on Majorana box & \eqref{Wopdef} \\ \hline
   $\omega_0$ & drive frequency & \eqref{Hdriv} \\ \hline
    $\omega_c$ & cut-off frequency for Ohmic bath & \eqref{spectraldensity}\\ \hline
     $T$ & temperature & \eqref{condit} \\ \hline
     $t_0$ & overall scale of tunnel couplings between QDs and Majorana box & \eqref{Htun} \\ \hline
    $t_{LR}$ & tunnel coupling connecting both Majorana boxes, see Sec.~\ref{sec4} & \eqref{LRCoup} \\ \hline \hline
\emph{Dynamical quantities:}& & \\ \hline
    $D$ & dark space dimension & Sec.~\ref{sec3e4}\\     \hline
    $\Delta_{z,x,y,m}$ & dissipative gap for the respective dark state & e.g., see \eqref{deltaz}\\ \hline
    $h_{\pm}, \tilde h^{}_\pm$ & Lamb shift parameters for full and reduced Lindblad eq., respectively & \eqref{rate2}, \eqref{mod22}\\ \hline
    $J_\pm, \Gamma_{\pm}, H_{\rm L}$ & jump operators, transition rates, and Hamiltonian for full Lindblad eq.  & \eqref{jumpops}, \eqref{rate2}, \eqref{jzdef}\\ \hline
    $\tilde J^{}_\pm, \tilde\Gamma^{}_{\pm}, \tilde H^{}_{\rm L}$ & jump operators, transition rates, and Hamiltonian for reduced Lindblad eq. & \eqref{jumpN}, \eqref{mod1}, \eqref{mod22}\\ \hline
      $K_{j=1,\ldots,6}, \tilde \Gamma^{}_{j}$ & jump operators and transition rates for two-box setup & \eqref{jumpK}, \eqref{trans22}\\ \hline
    $p$ & occupation probability of high-lying QD & \eqref{ssform}\\ \hline
     $\rho(t)$ & reduced density matrix for combined QD-Majorana system & \eqref{UnwantedTerms}\\ \hline
    $\rho_{\rm M}(t)$ & reduced density matrix for the Majorana sector & \eqref{factorize} \\ \hline
    $(\tau_x,\tau_y,\tau_z)$ & Pauli operators for QD pair in single-occupancy regime $N_{\rm d}=1$ & \eqref{taudef} \\ \hline
    $\theta_{j\nu},\theta$ & fluctuating electromagnetic phases & \eqref{hatlambda}, \eqref{phasedef}\\ \hline
     $\hat W_{jk}, \hat W_{x,y,z}$ & fluctuating cotunneling operators & \eqref{Wopdef}, \eqref{wdef1} \\ \hline
    $W_{jk}, W_{x,y,z}$  & cotunneling operators for $\theta_{j,\nu}=0$  & \eqref{wdef2}\\ \hline
    $(X,Y,Z)$ & Pauli operators of Majorana box & \eqref{PauliOp} \\ \hline
\hline
\end{tabular}
\caption{List of important symbols. 
\label{table1}}
\end{center}
\end{table*}

\section{Driven dissipative Majorana dynamics}\label{sec2}

We start this section by discussing the Majorana box \cite{Plugge2017,Karzig2017}.
Our DD model as well as the physical assumptions behind it are explained in Sec.~\ref{sec2a}. 
We then derive the Lindblad master equation governing the dynamics of the reduced density matrix of the Majorana sector.
To that end, we first trace over the environmental degrees of freedom in Sec.~\ref{sec2b},  and then over the QD fermions in Sec.~\ref{sec2c}.  

\subsection{Model and low-energy theory} \label{sec2a}

In this subsection, we introduce the model for the DD Majorana setup illustrated in Fig.~\ref{fig1}.  We also 
outline the hardware ingredients needed for implementing our dark state stabilization and manipulation protocols.  
 For concreteness, we refer to a possible realization using proximitized semiconductor nanowires \cite{Plugge2017,Karzig2017}.
In addition, we describe the effective low-energy Hamiltonian obtained
after the high-energy charge states on the Majorana island are projected away.

\subsubsection{Majorana box}

Consider the setup depicted in Fig.~\ref{fig1}, where a floating topological superconductor island harbors $M$ zero-energy MBSs. For this case we have $M=4$, but for generality, we shall allow for general (even) values of $M$.
The MBSs correspond to the Majorana operators $\gamma^{}_\nu=\gamma_\nu^\dagger$, with anticommutator 
$\{\gamma_\nu,\gamma_{\nu'}\}=2\delta_{\nu\nu'}$ and $\nu=1,\ldots,M$.  As indicated in Fig.~\ref{fig1}, they could be realized as end states of two parallel InAs/Al nanowires \cite{Lutchyn2018}.  We consider class-$D$ topological superconductor wires, where time reversal symmetry is broken by
a magnetic field \cite{Alicea2012}. 
Both nanowires are electrically connected by a superconducting bridge such that the entire island has a common charging energy, $E_C=e^2/(2C)$, with 
typical values of the order $E_C\approx 1$~meV \cite{Lutchyn2018}.  
The isolated island (`box') has the Hamiltonian (we work in the Schr\"odinger picture for now) 
\begin{equation} \label{Hbox}
    H_{\rm box}=E_C( \hat N-N_g)^2.
\end{equation}
The operator $\hat N$ refers to the total electron number on the box, and $N_g$ is a tunable backgate parameter. 
In Eq.~\eqref{Hbox} we have neglected 
hybridization energies resulting from a finite overlap between different MBS pairs. These energy scales are 
exponentially small in the respective MBS-MBS distance.  As will be discussed in Sec.~\ref{sec3e}, a small hybridization between MBSs 
is often tolerable for DD-generated dark states or dark spaces.  For the native Majorana qubit, such effects cause dephasing.

Our theory requires several conditions to be satisfied.  
First, we assume that our DD protocols only involve energy scales well below both $E_C$ and the superconducting (proximity) gap $\Delta$.
This assumption implies that the ambient temperature satisfies $T\ll {\rm min}\{E_C,\Delta\}$,  which typically requires temperatures below $100$~mK in semiconductor-based 
Majorana platforms \cite{Lutchyn2018}.
We can then neglect the effects of above-gap continuum quasiparticles, as has tacitly been assumed in Eq.~\eqref{Hbox}, which otherwise constitute an intrinsic source of dissipation in the Majorana sector.
In practice, one also needs to ensure that accidental low-energy Andreev states are not accessible, see Ref.~\cite{Manousakis2019} for a recent discussion. 
Second, we consider Coulomb valley conditions \cite{Nazarov,AltlandBook}, i.e.,  $N_g$ is tuned close to an integer value and the box is only weakly coupled to the QDs in Fig.~\ref{fig1}. In that case, $H_{\rm box}$ leads to  charge quantization, which dictates the fermion number parity of the island. At temperatures well below the superconducting gap,  
only the Majorana sector of the full Hilbert space of the box has to be kept \cite{Fu2010}. For $M=4$, 
we arrive at a parity constraint in the Majorana sector, $\gamma_1\gamma_2\gamma_3\gamma_4=\pm 1$ \cite{Beri2012}, and
the lowest-energy island state is then doubly degenerate. 
  The corresponding Pauli operators  associated with the resulting Majorana qubit
are represented by Majorana bilinears \cite{Beri2012,Landau2016,Plugge2016},
\begin{equation}\label{PauliOp}
X=i\gamma_1\gamma_3,\quad Y=i\gamma_3\gamma_2,\quad  Z=i\gamma_1\gamma_2.
\end{equation}
The fact that Pauli operators correspond to spatially separated pairs of Majorana operators allows for unusually versatile qubit access options.
The qubit is encoded either in the even parity sector, i.e., 
by using the two degenerate states with fermionic occupation number $N_m=0$ and $N_m=2$ in the Majorana sector (with one Cooper pair less for the $N_m=2$ state), 
or in the odd parity sector, where both states have $N_m=1$. 

\subsubsection{Quantum dots}

We next turn to the Hamiltonian describing the two QDs, $H_{\rm d}$,  in Fig.~\ref{fig1}. 
We start from a general single-dot Hamiltonian, $H_{\rm QD}=\sum_{\alpha} h_{\alpha} d_{\alpha}^{\dagger}d^{}_{\alpha}+\epsilon_C\left(\hat n-n_g\right)^2$,
where $\alpha$ labels electron spin and orbital degrees of freedom, $d_{\alpha}$ are fermion operators with $\hat n=\sum_{\alpha} d_{\alpha}^{\dagger}d^{}_{\alpha}$,  $h_{\alpha}$ describes a single-particle energy,
and $\epsilon_C$ is the (large) dot charging energy \cite{Karzig2017,Flensberg2011,Nazarov,AltlandBook}.   On low energy scales, 
the dot can then effectively be described by a single spinless fermion level. 
Denoting the corresponding level energy by $\epsilon_j$ for QD $j=1,2$, one arrives at
\begin{equation}\label{HDots}
 H_{\rm d}=\sum_{j=1,2} \epsilon_j d_j^{\dagger} d^{}_j,
\end{equation}
 see Ref.~\cite{Karzig2017} for details.
The energies $\epsilon_j$ can be controlled by variation of the gate voltage parameter $n_g$.  Without loss
of generality, we take $\epsilon_2>\epsilon_1$ throughout, where both energies  should
 satisfy $|\epsilon_j|\ll {\rm min}\{E_C,\Delta\}$.
In addition, we employ a time-dependent electromagnetic driving field which can pump single electrons between 
the two QDs via a tunnel link.
To that end, a suitable AC voltage can be applied to a  gate electrode located near this link. 
The respective Hamiltonian contribution is given by \cite{Platero2004}
\begin{equation}\label{Hdriv}
H_{{\rm drive}}(t) = w(t) d_1^{\dagger} d_2^{}+{\rm h.c.},\quad
w(t) = t_{12}+2A \cos\left(\omega_0 t\right),
\end{equation}
where $\omega_0$ denotes the drive frequency and $A$ the drive amplitude. In what follows, we assume that the static contribution vanishes, $t_{12}=0$, 
because a small coupling $t_{12}\ne 0$ will not affect the dissipator in the Lindblad equation, see Eq.~\eqref{GeneralLindblad} below,
and thus does not change the physics in a qualitative manner.

In this work, we  consider the Coulomb valley regime where the total charge on the box is fixed by the charging term in Eq.~\eqref{Hbox} on time scales $\delta t>1/E_C$ \cite{Romito2014}.
The total particle number on the QDs, $N_{\rm d}=\sum_j d_j^\dagger d_j^{}$, is therefore also conserved on such time scales.  For even $N_{\rm d}\in\{0,2\}$, the inter-QD dynamics is effectively frozen out.  We here mainly focus on the case $N_{\rm d}=1$, where the pair of QDs forms a spin-1/2 degree of freedom corresponding to  Pauli operators  
$\tau_{x,y,z}$ with $\tau_\pm=(\tau_x\pm i \tau_y)/2$,
\begin{equation}
\tau_+ = \tau_-^\dagger = d_1^{\dagger}d_2^{}, \quad
\tau_z = d_1^{\dagger}d_1^{} - d_2^\dagger d_2^{} = 2\tau_+\tau_--1.\label{taudef}
\end{equation}
We next turn to  the tunnel couplings connecting the QDs to the island.  

\subsubsection{Tunnel couplings and electromagnetic environment}

In the above parameter regime, tunneling to the box has to proceed via MBSs since no other low-energy island states are available. 
Such processes can be inelastic due to the coupling to a bosonic environment.  We here consider the case of a dissipative electromagnetic
environment, which can be modeled by including fluctuating phases $\theta_{j\nu}$ in the tunneling matrix elements \cite{Nazarov,Devoret1990,Girvin1990}, 
\begin{equation}\label{hatlambda}
\hat\lambda_{j\nu}=\lambda_{j\nu}e^{i\theta_{j\nu}},
\end{equation}
with dimensionless complex-valued parameters $\lambda_{j\nu}$ subject to ${\rm max}\{|\lambda_{j\nu}|\}=1$.
Here $\lambda_{j\nu}$ determines the transparency of the tunnel link between the QD fermion $d_j$ and the Majorana state $\gamma_\nu$ in the absence of electromagnetic noise
\cite{Zazunov2016}. 
The parameters $\lambda_{j\nu}$ play an important role in our DD scheme below.  Both their amplitude as well as their phase can be tuned by varying the voltage on 
a local gate near the tunnel contact in question, see Ref.~\cite{Lutchyn2018} and references therein.
With the overall hybridization energy $t_0$ characterizing 
the QD-MBS couplings,  the tunneling Hamiltonian is then given by \cite{Nazarov,Devoret1990,Girvin1990}
\begin{equation}\label{Htun}
 H_{{\rm tun}} =t_0  e^{-i\hat\phi}\sum_{j,\nu} \hat \lambda_{j\nu} d^{\dagger}_j \gamma_\nu  + {\rm h.c.}
\end{equation}
The phase operator $\hat\phi$ of the island has the commutator $[\hat N,\hat\phi]=-i$ with the number operator $\hat N$  in Eq.~\eqref{Hbox}.
The $e^{i\hat\phi}$  ($e^{-i\hat \phi}$) factor in Eq.~\eqref{Htun} thus ensures that an electron charge is added to (subtracted from) the island in a tunneling process.
It is well known that the electromagnetic potential fluctuations predominantly couple to the phase of the wave function \cite{Devoret1990,Girvin1990}. This fact is
expressed by the appearance of the fluctuating tunnel couplings $\hat\lambda_{j\nu}$, see
Eq.~\eqref{hatlambda}, in the tunneling Hamiltonian \eqref{Htun}.

For concreteness, we assume that the electromagnetic environment can be modeled by a single bosonic bath, see also Ref.~\cite{Munk2019}.
Representing the bath by an infinite set of harmonic oscillators \cite{Weiss2007,Breuer2006},  the environmental Hamiltonian is 
$H_{\rm env}=\sum_m E_m b_m^{\dagger} b_m^{}$, with the  energy $E_m>0$ of the photon mode
described by the boson annihilation operator $b_m$. In practice, the relevant bath energies $E_m$ are  strongly suppressed above a cutoff frequency 
$\omega_c$.  With dimensionless real-valued couplings $g_{j\nu,m}$, the stochastic phase operators $\theta_{j\nu}$ are  written as 
\begin{equation}\label{phasedef}
    \theta_{j\nu}=\sum_{m} g_{j\nu,m} \left(b^{}_m+b_m^\dagger\right).
\end{equation}
Clearly, they commute with each other, $[\theta_{j\nu},\theta_{j'\nu'}]=0$. 

\subsubsection{Low-energy theory}\label{spsec1}

We are interested in the parameter regime defined by the conditions
\begin{equation}\label{condit}
    {\rm max}\{T,A,t_0,\omega_0,\omega_c,|\epsilon_j|\}\ll {\rm min}\{E_C,\Delta\}.
\end{equation}
The parameters on the left side of Eq.~\eqref{condit} affect the dissipative transition rates in the Lindblad equation \eqref{GeneralLindblad} below. These rates in turn
set the time scale on which dark states are approached. 
We will adopt  a concise description, whereby for engineering a stabilization protocol targeting a specific dark state, it suffices to adjust the complex-valued tunnel link parameters $\lambda_{j\nu}$,
see Sec.~\ref{sec3}.   In practice, those \emph{state design parameters} can be changed via gate voltages.  
We also note that under the conditions in Eq.~(\ref{condit}), boson-assisted processes can neither excite above-gap quasi-particles nor higher-energy charge states on the island.

The full Hamiltonian can then be projected onto the doubly degenerate ground-state space of the box, $H(t)\to H_{\rm eff}(t)$. 
Using a Schrieffer-Wolff transformation to implement this projection, and noting that $H_{\rm box}$ then reduces to an irrelevant constant energy shift, 
we arrive at the effective low-energy Hamiltonian
\begin{equation}\label{heff0}
H_{\rm eff}(t)=H_{\rm d}+ H_{\rm env} + H_{\rm drive}(t)+H_{\rm cot},
\end{equation}
with the drive term in Eq.~\eqref{Hdriv} and the cotunneling contribution
\begin{equation}
 H_{\rm cot} = g_0\sum_{j,k=1,2}\hat W_{jk} 
 \left( 2 d_j^\dagger d_k^{}-\delta_{jk} \right), \quad g_0\equiv \frac{t_0^2}{E_C}.\label{Heff}
 \end{equation}
We here use the operators 
 \begin{equation}\label{Wopdef}
 \hat W_{jk}= \sum_{1\le\mu<\nu\le M} \left( 
 \hat \lambda_{j\nu} \hat \lambda_{k\mu}^\dagger
 - \hat \lambda_{j\mu}\hat \lambda_{k\nu}^\dagger \right)\gamma_\mu\gamma_\nu.
\end{equation} 
Equation \eqref{Heff} describes  cotunneling paths through the box, where the energy of the intermediate virtual
state has been approximated by $E_C$, cf.~Eq.~\eqref{condit}, and photon emission and absorption processes are encoded by the $\hat \lambda$ factors in Eq.~\eqref{hatlambda}.

For even QD occupation number $N_{\rm d}$, Eq.~\eqref{Heff} reduces to 
\begin{equation}\label{heffeven}
    H_{\rm cot}^{(N_{\rm d}=0,2)}=g_0 \ {\rm sgn}(N_{\rm d}-1) \ \sum_j \hat W_{jj}.
\end{equation}
For $N_{\rm d}=1$, using the notation 
\begin{eqnarray}
\nonumber
    \hat W_+&\equiv& \hat W_{12}, \quad \hat W_-=\hat W_+^\dagger,\quad \hat W_x=\hat W_++\hat W_-,\\ \label{wdef1}
     \hat W_y&=&i(\hat W_+ -\hat W_-),\quad 
\hat W_z=\hat W_{11}-\hat W_{22}, 
\end{eqnarray}
we find that Eq.~\eqref{Heff} can instead be expressed in the form 
\begin{equation}\label{cot1}
    H_{\rm cot}^{(N_{\rm d}=1)} = g_0 \sum_{a=x,y,z} \hat W_a \tau_a, 
\end{equation}
with the QD Pauli operators $\tau_a$ in Eq.~\eqref{taudef}.
We emphasize that like the $\hat W_{jk}$ operators in Eq.~\eqref{Wopdef}, also the $\hat W_a$ still contain the phase fluctuation operators 
due to the electromagnetic environment.
In order to realize the most general qubit-qubit exchange coupling between the QD spin $\{\tau_a\}$ and the $M=4$ Majorana box spin $(X,Y,Z)$ 
in the cotunneling regime, one has to specify nine independent (tunable) real-valued coupling constants.  
For the $M=4$ case in Fig.~\ref{fig1}, taking into account gauge invariance --- which allows us to set one of the $\lambda_{j\nu}$ to a real value ---, the five different
complex-valued hopping parameters $\lambda_{j \nu}$ are sufficient.
On top of that, the direct tunnel amplitude between the QDs is assumed to be real-valued after setting $t_{12} = 0$ in Eq.~\eqref{Hdriv}.

To simplify the subsequent analysis, we assume that the dominant contribution to the environmental electromagnetic noise comes from the long wavelength part.
In effect, such contributions will cause dephasing of the QDs, e.g., due to the presence of a backgate electrode.  This assumption is also consistent with the picture of a single bath. To good accuracy, the couplings $g_{j\nu,m}$ in Eq.~\eqref{phasedef} then do not depend on the Majorana index $\nu$, i.e., $g_{j\nu,m}=g_{j,m}$.
As a consequence, also the fluctuating phases \eqref{phasedef} become $\nu$-independent, $\theta_{j\nu}=\theta_j$.  In that case, the diagonal entries $\hat W_{jj}$ 
are insensitive to electromagnetic noise and
the bath completely decouples  for even $N_{\rm d}$, see Eq.~\eqref{heffeven}.

From now on, we therefore focus on the case of a single electron shared by the QDs, $N_{\rm d}=1$. Defining the phase operator
\begin{equation}
    \theta\equiv \theta_1-\theta_2 =  \sum_m (g_{1,m}-g_{2,m}) \left(b_m^{}+b_m^\dagger\right),
\end{equation} 
Eq.~\eqref{cot1} then yields 
\begin{equation}\label{cotfinal}
    H_{\rm cot}= 2g_0 \left(e^{i\theta} W_+\tau_+ + {\rm h.c.}\right)+ g_0 W_z \tau_z.
\end{equation}
The operators $W_+$ and $W_z$ correspond to `undressed' ($\theta_{j\nu}\to 0$) versions of $\hat W_+$ and $\hat W_z$, respectively.
These operators act only on the Hilbert space sector describing Majorana states. Comparing to Eq.~\eqref{Wopdef}, we have
\begin{equation}\label{wdef2}
W_{jk} = \sum_{\mu<\nu}^M \left(   \lambda_{j\nu}  \lambda_{k\mu}^\ast
 -  \lambda_{j\mu} \lambda_{k\nu}^\ast \right)\gamma_\mu\gamma_\nu.
\end{equation}
For the device in Fig.~\ref{fig1}, the $W_{jk}$ operators can be expressed in terms of the Pauli operators $(X,Y,Z)$ in Eq.~\eqref{PauliOp}, see below.

\subsubsection{Bath correlation functions}

The equilibrium density matrix of the thermal environment is given by
\begin{equation}\label{envtrace}
\rho_{\rm env}=Z^{-1}_{\rm env} e^{-H_{\rm env}/T}\quad {\rm with} \quad Z_{\rm env}={\rm tr}_{\rm env} \ e^{-H_{\rm env}/T},
\end{equation}
with `tr$_{\rm env}$' denoting a trace operation over the environmental bosons.
Using $\braket{\hat O}_{\rm env}\equiv {\rm tr}_{\rm env} (\hat O\rho_{\rm env})$, 
we define the  correlation function  \cite{Weiss2007}
\begin{eqnarray} \nonumber
&& J_{\rm env}(t)=\braket{[\theta(t)-\theta(0)] \theta(0)}_{\rm env} = 
\int_0^\infty \frac{d\omega}{\pi} \frac{{\cal J}(\omega)}{\omega^2}\times  \\
 &&\quad \times \left\{ [\cos(\omega t)-1] \coth\left(\frac{\omega}{2T}\right)  - i \sin(\omega t) \right\},\label{BathCorr}
\end{eqnarray}
with the spectral density 
 \begin{equation}\label{spectraldensity}
 {\cal J}(\omega)=\pi\sum_m (g_{1,m}-g_{2,m})^2 E_m^2 \delta(\omega-E_m).
\end{equation}
Switching to the continuum limit in bath frequency space, we  focus on the practically most important Ohmic case with ${\cal J}(\omega)\propto \omega$ in the low-frequency limit. 
In concrete calculations, we use the model spectral density  \cite{Weiss2007} 
\begin{equation}\label{Ohmic}
 {\cal J}(\omega) = \alpha \omega e^{-\omega/\omega_c},
\end{equation}
where $\alpha$ is a dimensionless system-bath coupling and frequencies above $\omega_c$ are exponentially suppressed.  For a related discussion in the context of Majorana qubits, see Ref.~\cite{Munk2019}.  The parameter $\alpha$ is related to the environmental impedance $Z(\omega)$ \cite{Devoret1990},
\begin{equation}\label{alphadef}
    \alpha= \frac{e^2}{2h}  {\rm Re} Z(\omega=0).
\end{equation}
We consider the case $\alpha<1$ below.

For the subsequent discussion, we rewrite $H_{\rm cot}$ in normal-ordered form relative to the phase fluctuations, 
\begin{equation}\label{normalorder1}
    H_{\rm cot} = H_{\rm cot}^{(0)} + V,
\end{equation}
where $H_{\rm cot}^{(0)}$ is the expectation value of $H_{\rm cot}$ with respect to phase fluctuations and $V$ represents the coupling of the combined QD-MBS system to
phase fluctuations.
Since $\langle\theta^2\rangle_{\rm env}$ diverges in the Ohmic case, we have $\langle e^{i\theta}\rangle_{\rm env}=0$, resulting in 
\begin{equation}\label{normalorder2}
    H_{\rm cot}^{(0)}  \equiv  \braket{H_{\rm cot}}_{\rm env}= g_0 W_z\tau_z.
\end{equation}
The interaction term in Eq.~\eqref{normalorder1} is then given by
\begin{equation}\label{normalorder3}
    V= 2g_0 \left( e^{ i\theta} W_+ \tau_+ + {\rm h.c.}\right) .
\end{equation}
By construction, $\braket{V}_{\rm env}=0$. 
Correlation functions of exponentiated phase fluctuations are given by ($s=\pm 1$)
\begin{equation}\label{xicor}
  \braket{e^{is\theta (t)} e^{-is\theta(0)}}_{\rm env} =  e^{J_{\rm env}(t)}  
\end{equation}
with $J_{\rm env}(t)$ in Eq.~\eqref{BathCorr}.  

\subsubsection{Interaction picture and Rotating Wave Approximation}

From now on, we shall switch to the interaction picture with respect to $H_{\rm d}+H_{\rm env}$. 
The Hamiltonian then takes the form,
see~Eqs.~\eqref{heff0} and \eqref{normalorder1},
\begin{eqnarray}\label{Hfull}
 H_{{\rm eff},I}(t) &=&  H_{0,I}(t)+  V_I(t),\\ \nonumber H_{0,I}(t) &=& H_{{\rm drive},I}(t) + H^{(0)}_{{\rm cot},I}(t).
\end{eqnarray}
For simplicity, we drop the `$I$' index (for interaction picture) in what follows and focus on resonant drive conditions, 
\begin{equation}\label{omegadef}
    \omega_0=\epsilon_2-\epsilon_1.
\end{equation}
In the regime $\omega_0\gg T$ considered below, see Eq.~\eqref{basiccond}, 
we can then apply the rotating wave approximation (RWA) \cite{Gardiner2004}.
As a consequence, $H_{\rm drive}(t)\to \tilde H_{\rm drive}$ with
\begin{equation}\label{Hdriv2}
 \tilde H_{\rm drive}=A\left( d_1^{\dagger} d^{}_2+ d_2^{\dagger} d^{}_1\right) = A \tau_x,
\end{equation}
resulting in a time-independent drive Hamiltonian in the interaction picture.
If the drive frequency is slightly detuned,
$\omega_0=\epsilon_2-\epsilon_1+ \delta\omega_0$,
a residual time dependence remains, $H_{\rm drive}(t)= e^{-i\delta\omega_0 t} A  d_1^{\dagger} d^{}_2 +$~h.c., after applying the RWA.
However, we find that the final Lindblad equation for the dynamics of the Majorana sector in Sec.~\ref{sec2c} 
is not affected to leading order in $\delta\omega_0$.  A small mismatch in the resonance condition \eqref{omegadef} will therefore not obstruct our findings.  We then put $\delta\omega_0=0$ from now on.

\subsection{Master equation}\label{sec2b}

In this subsection, we consider the time evolution of the reduced density matrix, $\rho(t)$, describing the coupled system defined by the MBSs and the pair of QD fermions. 
After tracing over the environmental bosons, we arrive at a Lindblad master equation for the dynamics of $\rho(t)$.
In Sec.~\ref{sec2c}, we will subsequently trace over the QD fermions to obtain a Lindblad equation for the Majorana sector only.
With $\omega_0=\epsilon_2-\epsilon_1$ and $g_0=t_0^2/E_C$, we  consider the regime
\begin{equation}\label{basiccond}
   g_0\ll T\ll  \omega_0, \quad A\alt g_0.
\end{equation} 
In particular, $T\ll \omega_0$ is needed to justify the RWA, while $g_0\ll T$ is required for the Born-Markov approximation. 
In addition, the regime $g_0\ll T$ allows us to neglect emission and absorption processes taking place only in the Majorana sector since the bath is then unable to resolve such transitions.  Of course, we will account for boson-assisted inter-QD transitions resulting from cotunneling processes. 
Equation~\eqref{basiccond} also states that we study a weakly driven system with drive amplitude $A\alt g_0$.  The opposite strongly driven case is briefly discussed in App.~\ref{appA}.
We note that inelastic corrections to the drive Hamiltonian due to electromagnetic phase fluctuations, see Eq.~\eqref{envtrace}, can be neglected by the secular approximation, cf. Sec.~II.B of Ref.~\cite{Shavit2019}.
We show below that the parameters appearing in Eq.~\eqref{basiccond} will only affect the speed of approach towards the targeted dark state (or dark space) but not the state fidelity. 
Moreover, our protocols turn out to be exceptionally robust under even $10\%$ mismatch in \emph{all} tunneling amplitudes which in turn may affect the state fidelity, see, e.g., 
Fig.~\ref{fig5} below.  We therefore expect that, in practice, it is not necessary to fulfill the `$\ll$' inequalities in Eq.~\eqref{basiccond} in an overly strict sense.

\subsubsection{Lindblad master equation for $\rho(t)$}

The master equation governing the dynamics of the density matrix $\rho(t)$ for the combined system (QDs and Majorana sector) is obtained by following the standard derivation of 
Born-Markov master equations \cite{Weiss2007,Breuer2006,Gardiner2004}.   
We assume a factorized initial (time $t=0$) density matrix of the total system,
$\rho_{\rm tot}(0)=\rho(0) \otimes \rho_{\rm env}$, with $\rho_{\rm env}$ in Eq.~\eqref{envtrace}.
Starting from the von-Neumann equation for $\rho_{\rm tot}(t)$ subject to $H_{\rm eff}(t)$ in Eq.~\eqref{Hfull}, 
we trace over the environmental modes and apply the Born-Markov approximation \cite{Weiss2007,Breuer2006,Gardiner2004}.
As a result, $\rho(t)$ obeys the master equation
\begin{eqnarray} \label{UnwantedTerms}
&&\partial_t\rho(t)=-i\left[ H_{0}(t),\rho(t)\right] -
{\rm tr}_{\rm env}
\int_0^{\infty} d\tau \\ \nonumber &\times& \left[ V(t),\left[ V(t-\tau)+H_0(t-\tau),\rho(t)\otimes\rho_{\rm env}\right]\right], 
\end{eqnarray}
where we have used that, by construction, ${\rm tr}_{\rm env}\left[ V(t),\rho(0)\otimes\rho_{\rm env}\right]=0$.  
Similarly, the mixed term involving $V(t)$ and $H_{0}(t-\tau)$ vanishes identically.  We are 
left with the coherent evolution term due to $H_{0}$ and the double commutator containing two $V$ terms.  

Unfolding the double commutator, 
we arrive at a master equation of Lindblad \cite{Lindblad1976,Lindblad1983} type,  
\begin{equation}\label{GeneralLindblad}
\partial_t\rho(t) = -i\left[ H_{\rm L},\rho(t)\right]+  \sum_{\pm} \Gamma_\pm  \mathcal{L}[J_\pm]\rho(t) .
\end{equation}
The subscript `L' in $H_{\rm L}$ is meant to clarify that this Hamiltonian appears in a Lindblad equation.
The dissipator ${\cal L}$ acts as  superoperator on $\rho$ \cite{Breuer2006},
\begin{equation}\label{Dissipator}
    \mathcal{L}[J]\rho=J\rho J^\dagger -\frac{1}{2}\lbrace J^{\dagger} J,\rho \rbrace.
\end{equation}
The two \emph{jump operators} in Eq.~\eqref{GeneralLindblad} are given by
\begin{equation}\label{jumpops}
    J_\pm = 2W_\pm \tau_\pm = J_\mp^\dagger,
\end{equation}
with the corresponding dissipative transition rates,
\begin{equation}\label{dissrate}
    \Gamma_{\pm}= 2g_0^2\, {\rm Re} \Lambda_\pm  .
\end{equation}
Here, we define the quantities
\begin{equation}\label{lambdadef}
    \Lambda_\pm= \int_0^\infty dt \, e^{\pm i\omega_0  t} e^{J_{\rm env}(t)},
\end{equation}
with the bath correlation function \eqref{BathCorr}.
Their imaginary parts give Lamb shift parameters,
\begin{equation}\label{shifts2}
    h_\pm = g_0^2 \, {\rm Im}\Lambda_\pm ,
\end{equation}
which appear in the Hamiltonian governing
the coherent time evolution in Eq.~\eqref{GeneralLindblad},
\begin{equation}\label{Hq}
    H_{\rm L} = A \tau_x + g_0 W_z\tau_z + \sum_{\pm}  h_\pm  J_{\pm}^{\dagger} J_\pm^{}.
\end{equation}
The first two terms in $H_{\rm L}$ originate from $H_0$ in Eq.~\eqref{Hdriv2}, while
the third term contains the Lamb shifts \eqref{shifts2}.

Next we observe that Eq.~\eqref{BathCorr} implies the general relation 
\begin{equation}\label{detbal}
    J_{\rm env}\left(-t-i/T\right)=J_{\rm env}(t)
\end{equation}
in the complex-time plane. 
Using Eq.~\eqref{detbal} in Eq.~\eqref{lambdadef} then results in a detailed balance relation, $\Lambda_-=e^{-\omega_0/T} \Lambda_+$.  As a consequence, for arbitrary parameters, we find
\begin{equation}
\Gamma_-=e^{-\omega_0/T}\Gamma_+,  \quad h_-=e^{-\omega_0/T} h_+.
\end{equation} 
In particular, for $T\ll \omega_0$, the dissipative rate $\Gamma_-$ associated with the jump operator $J_-$ will be
exponentially suppressed against the rate $\Gamma_+$.   The dissipative  part of the Lindblad equation 
\eqref{GeneralLindblad} is therefore completely dominated by the jump operator $J_+$.

It is a distinguishing feature of our DD platform
that  jump operators can be directly implemented by designing \emph{unidirectional} inelastic cotunneling paths connecting pairs of QDs via the box, with the overall energy scale $g_0$. 
The QDs are also directly coupled by a driven tunnel link $w(t)$, see Eq.~\eqref{Hdriv}, with overall energy scale $A$.  For $T\ll \omega_0$, as far as inter-dot transitions via the box are concerned, only photon emission processes are relevant. As a consequence,  only  transitions from the energetically high-lying QD 2 to QD 1 may take place, corresponding to the jump operator  $J_+\propto \tau_+$, see Eqs.~\eqref{taudef} and \eqref{jumpops}.  
Such transitions act on the Majorana state according to the operator $W_+$.  As we show below, this operator can be engineered at will by adjusting the tunneling parameters $\lambda_{j\nu}$, which in turn is possible by changing suitable gate voltages.  The driving field pumps the dot electron in the opposite direction, i.e., from QD $1\to 2$, and for a small pumping rate, $A\alt g_0$, we obtain a 
steady state circulation $1\to 2\to 1$ by alternating pumping and cotunneling processes.  On the other hand, for $A>g_0$, pumping processes will dominate and the cotunneling channel is effectively suppressed, see App.~\ref{appA}.

To facilitate analytical progress,
we consider the case $\omega_0\ll \omega_c$. (Otherwise Eq.~\eqref{lambdadef} can be solved numerically in a straightforward manner.)  One then finds \cite{Weiss2007}
\begin{equation}
    J_{\rm env}(t) \simeq -2\alpha \ln\left( \frac{\omega_c}{\pi T} \sinh(\pi T t)\right) - i\pi\alpha \,{\rm sgn}(t),
\end{equation}
and with the Gamma function $\Gamma(z)$, we arrive at
\begin{eqnarray}
\label{rate2}   
\Gamma_+ &\simeq& \Gamma(1-2\alpha)\sin(2\pi\alpha) \left(\frac{\omega_0}{\omega_c}\right)^{2\alpha} \frac{2 g_0^2 }{\omega_0} ,\\
\nonumber
    h_+ &\simeq& \frac12 \cot(2\pi \alpha) \Gamma_+.
\end{eqnarray}
 
For the device in Fig.~\ref{fig1}, using the Pauli operators \eqref{PauliOp}, 
 the jump operators $J_\pm^{}=J_\mp^\dagger$ follow
from Eq.~\eqref{jumpops} in the general form 
\begin{eqnarray}\nonumber
J_{+}&=& \tilde J_+ \tau_+ , \\
\tilde J_+ &=& 2ie^{i\beta_2}|\lambda_{23}|\left(e^{-i\beta_3}|\lambda_{11}|X-e^{-i\beta_1}|\lambda_{12}|Y\right) \nonumber
\\ &-& 2i\left[e^{-i\beta_1}|\lambda_{12}\lambda_{21}|-e^{i\beta_4}|\lambda_{11}\lambda_{22}|\right]Z, \label{jumpN}
\end{eqnarray}
where the phases $\beta_{1,2,3,4}$ are indicated in Fig.~\ref{fig1}. 
They are connected to the phases $\chi_{j\nu}$ in the tunneling parameters, $\lambda_{j\nu}=|\lambda_{j\nu}|e^{-i\chi_{j\nu}}$, by 
the relations 
\begin{equation}\label{betadef}
    \beta_1 = \chi_{12}, \quad \beta_2 = \chi_{23},\quad \beta_3= \chi_{11}, \quad \beta_4=\chi_{22},
\end{equation}
with the gauge choice $\chi_{21}=0$.
In particular,  $\beta_1-\beta_3$ ($\beta_2$) is the loop phase accumulated along the shortest closed tunneling trajectory involving only QD 1 (QD 2), cf.~Eq.~\eqref{jzdef}.  
These phases, as well as the absolute values $|\lambda_{j\nu}|$, can be experimentally varied, e.g., by changing the voltages on nearby gates. 
We emphasize that $\tilde J_+$ is fully determined by selecting the state design parameters $\lambda_{j\nu}$. 
The Hamiltonian $H_{\rm L}$ then follows as
\begin{eqnarray}
    H_{\rm L}&=& A\tau_x + 2g_0 \tilde J_z \tau_z + \sum_\pm h_\pm J_\pm^\dagger J_\pm^{},\nonumber \\  \label{jzdef}
   \tilde J_z &=& \frac12\bar\lambda^2+\sin\beta_2|\lambda_{21}\lambda_{23}| X+\\ \nonumber
    &+& \sin\left(\beta_4-\beta_2\right)|\lambda_{22}\lambda_{23}|Y+\\ \nonumber
&+&\left[\sin\beta_4|\lambda_{21}\lambda_{22}|-\sin\left(\beta_1-\beta_3\right)|\lambda_{11}\lambda_{12}|\right]Z  ,
\end{eqnarray}
where $\bar \lambda^2\equiv |\lambda_{11}|^2+|\lambda_{12}|^2+|\lambda_{21}|^2+|\lambda_{22}|^2+|\lambda_{23}|^2$. 
It is worth mentioning that the operators $\tilde J_\pm$ and $\tilde J_z$ act only on the Majorana subsector.

To illustrate the above general expressions, let us consider a simple example.  We take stabilization parameters subject to the conditions
\begin{eqnarray}\label{parameterchoice1}
    |\lambda_{11}| &=& |\lambda_{12}|, \quad \lambda_{22}=0, \\ 
    \nonumber \beta_1 &=&- \beta_2 =\pi/2, \quad \beta_3 = \beta_4 =0.
\end{eqnarray}
Using Eq.~\eqref{jumpN}, the dominant jump operator contributing to the Lindbladian is then given by 
\begin{equation}
 J_+=2|\lambda_{11}| \left(2|\lambda_{23}|\sigma_+ + |\lambda_{21}|Z\right)\tau_+, 
 \label{NumericJumpOperator}
\end{equation}
where $\sigma_\pm=(X\pm iY)/2$.  For $|\lambda_{23}|\gg |\lambda_{21}|$, the Lindbladian will then automatically drive an arbitrary Majorana state $\rho_{\rm M}$ towards $|0\rangle\langle 0|$, with the 
$Z$-eigenstate $|0\rangle$ to eigenvalue $+1$, i.e., $Z|0\rangle=|0\rangle$.  Here, the reduced density matrix $\rho_{\rm M}(t)$ describes the Majorana sector only, see Sec.~\ref{sec2c}.
However, the operator $\tilde J_z$ appearing in the Hamiltonian $H_{\rm L}$ still contains a small $X$ component, see Eq.~\eqref{jzdef}, which could potentially disrupt the action of the dissipator. 
Nonetheless, we find below that for small $|\lambda_{21}|$, the desired state $|0\rangle$ is approached with high fidelity,  regardless of the initial system state $\rho(0)$.
An optimized parameter choice for stabilizing  $|0\rangle$ will be discussed in Sec.~\ref{sec3}.

\subsubsection{Numerical results}\label{specialsec}

\begin{figure}[t]
\begin{centering}
\includegraphics[width=\columnwidth]{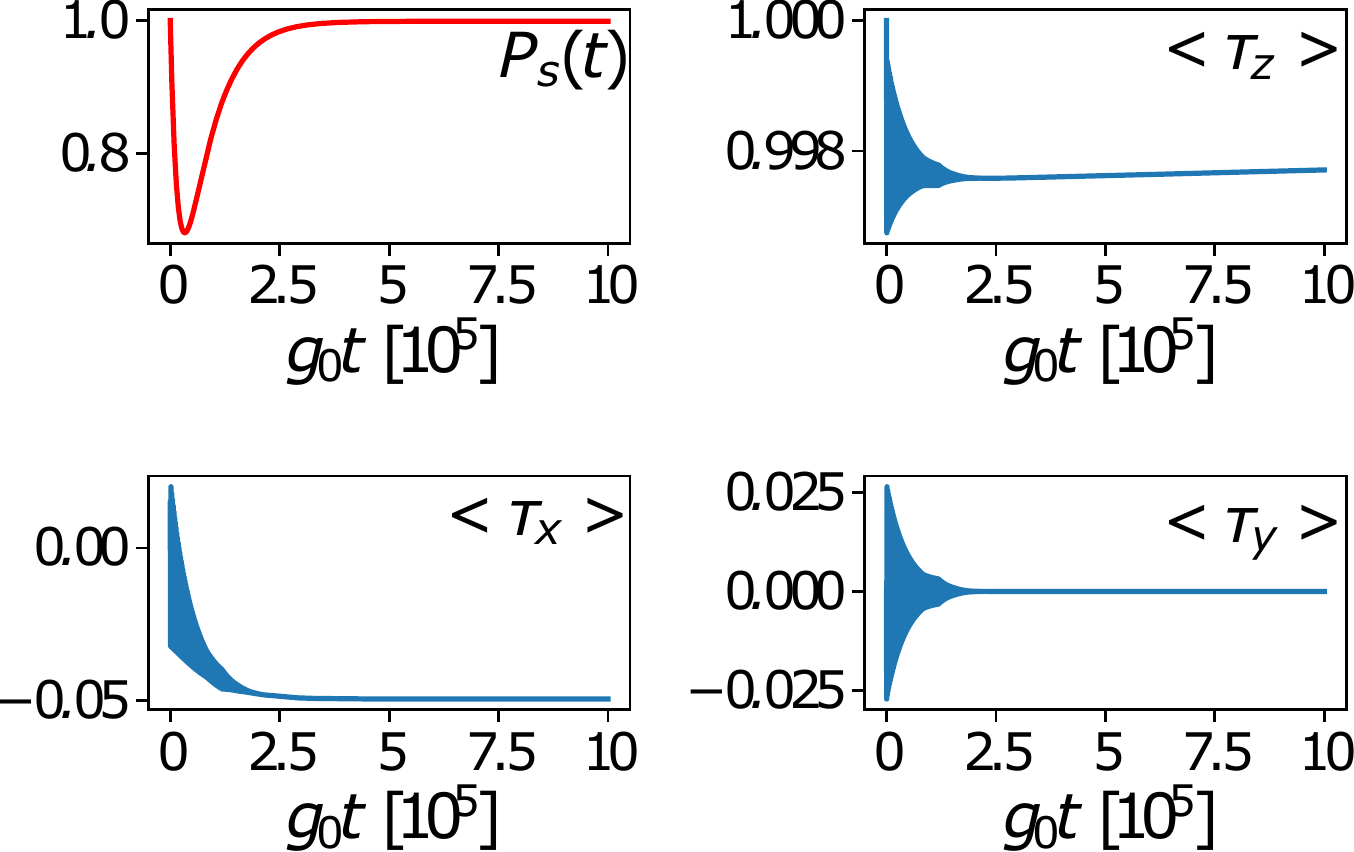}
\end{centering}
\caption{ Driven dissipative dynamics for the setup in Fig.~\ref{fig1}, illustrating
the time-dependent expectation values of the Pauli operators $\tau_{x,y,z}$ describing the QDs, see Eq.~\eqref{taudef}. We also show the 
purity, $P_s(t)$, of the system state, see Eq.~\eqref{puritydef}. All results were obtained by numerical integration of the Lindblad equation \eqref{GeneralLindblad} for the density matrix $\rho$ describing the QDs and the Majorana sector, with $H_{\rm L}$ in Eq.~\eqref{jzdef}.
We used the parameters  in Eq.~\eqref{parameterchoice1}, with $T/g_0=4$, $\omega_0/g_0=40$, $\omega_c/g_0=200$, $A/g_0=0.1$, $\alpha=1/4$, $|\lambda_{11}|=|\lambda_{12}|=|\lambda_{23}|= 1$, and $|\lambda_{21}|=0.1$.  Fast transient oscillations in $\langle\tau_a(t)\rangle$ are not resolved on the shown time scale, corresponding to shaded regions.
The respective dynamics in the Majorana sector is depicted in Fig.~\ref{fig3}. }
\label{fig2}
\end{figure}

\begin{figure}[t]
\begin{centering}
\includegraphics[width=0.7\columnwidth]{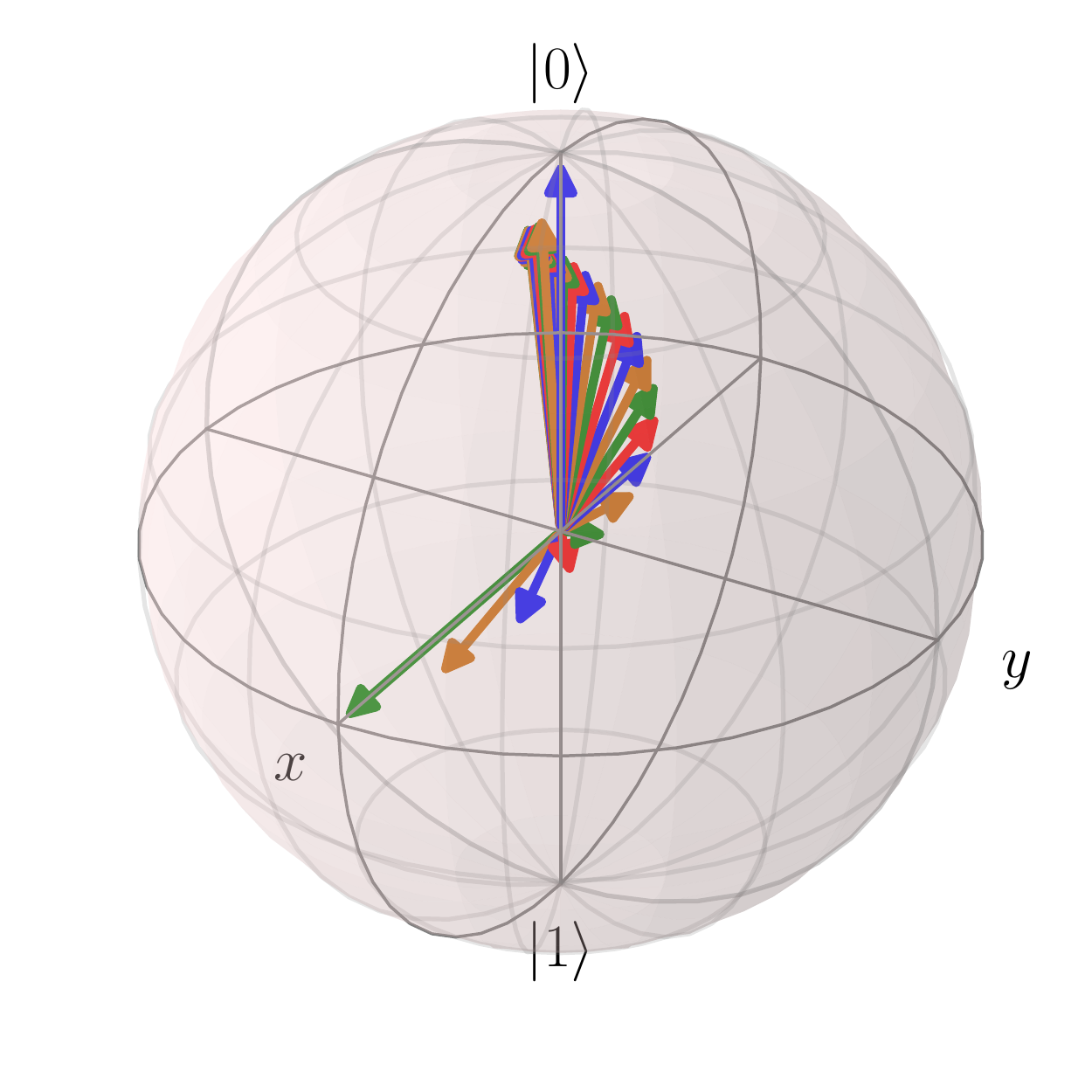}
\end{centering}
\caption{Time evolution of the Bloch vector, $(\langle X\rangle,\langle Y\rangle,\langle Z\rangle)(t)$, describing the Majorana state $\rho_{\rm M}(t)$ for the same parameters as in Fig.~\ref{fig2}.  The expectation value is computed by numerically integrating the Lindblad equation. Starting from the initial $X$-eigenstate $|+\rangle$, the DD protocol stabilizes the dark state $|0\rangle$ at long times, corresponding to the north pole of the Bloch sphere. The intermediate states (with alternating colors) were obtained at times $g_0t\in\lbrace 5\times 10^{3}, 10\times 10^3,\ldots,15\times10^4\rbrace$.  
\label{fig3}}
\end{figure}

We next turn to a numerical integration of Eq.~\eqref{GeneralLindblad} using the approach of Refs.~\cite{Johansson2012,Johansson2013}.
Numerical results for the above parameters are shown in Figs.~\ref{fig2} and \ref{fig3}. While the goal of the DD protocol is to stabilize a selected state in the Majorana sector, 
it is useful to also study the dynamics in the QD sector, see Fig.~\ref{fig2}.
We start from a pure initial state, $\rho(0)=|\Psi(0)\rangle\langle\Psi(0)|$, with $|\Psi(0)\rangle= |+\rangle \otimes |0\rangle_{\rm d}$,
where the $\tau_z=+1$ QD eigenstate, $|0\rangle_{\rm d}$, describes an electron located in the energetically lower QD 1, with 
QD 2 left empty, see Eq.~\eqref{taudef}. The initial Majorana state has been chosen as the $X$-eigenstate $|+\rangle$ with 
eigenvalue $+1$.  However, we have checked that the same long-time limit of $\rho(t)$ is reached for other initial states.
We define the purity of the system state as
\begin{equation}\label{puritydef}
P_s(t)= {\rm tr} \rho^2(t).
\end{equation}
The upper left panel of Fig.~\ref{fig2} shows that the purity approaches a value close to the largest possible value ($P_s=1$) at long times.  Moreover, as observed from Fig.~\ref{fig3}, the DD protocol steers the Majorana state towards the pure state $|0\rangle$, i.e., towards the north pole of the corresponding Bloch sphere. For the shown example, the QD state $\rho_{\rm d}$ has most probability weight in the energetically lower QD 1. Indeed, Fig.~\ref{fig2} shows that at long times, the electron shared by the two QDs will predominantly relax to QD 1, 
corresponding to the state $|0\rangle_{\rm d}$.   
Nonetheless, it is of crucial importance that the occupation probability $p$ for encountering the electron in the energetically higher QD 2 (corresponding to the state $|1\rangle_{\rm d}$) remains finite at long times.   
We find $p\approx 0.001$ for the parameters in Fig.~\ref{fig2}.  

We conclude that
the system state factorizes at long times, $\rho(t)\simeq \rho_{\rm M}\otimes \rho_{\rm d}$ with $\rho_{\rm M}=|0\rangle \langle 0|$.  
The approach of the Majorana state towards $|0\rangle$ takes place on a time scale given by the inverse of the dissipative gap of the reduced Lindbladian describing the Majorana sector only, see Sec.~\ref{sec3} below. The relaxation time scales for the QD subsystem can be longer, cp.~Figs.~\ref{fig2} and \ref{fig3}. 
 
Finally, we remark that for the special case $\lambda_{21}=0$, the electron shared by the two QDs will \emph{not} predominantly relax to the energetically lower QD $1$. One here has only two cotunneling paths between both QDs, namely the constituents forming the operator $4|\lambda_{11}\lambda_{23}|\sigma_+$ in Eq.~\eqref{NumericJumpOperator}. Both paths  interfere destructively once the Majorana island is stabilized in the state $\ket{0}$. An arbitrarily weak drive can then overcome all dissipative effects in the long-time limit.
In contrast to what happens for $\lambda_{21}\ne 0$, the QDs will thus realize an 
equal-weight mixture of $|0\rangle_{\rm d}$ and $|1\rangle_{\rm d}$. 
Nonetheless, the reduced Lindblad equation \eqref{LindbladMBQ} below still applies, with $p\to 1/2$ and $p_\perp\to 0$ in Eq.~\eqref{ssform}. We note that those parameters are also appropriate in 
the strongly driven case, cf.~App.~\ref{appA}.

\subsection{Lindblad equation for the Majorana sector} \label{sec2c}

The above observations allow us to derive a reduced Lindblad equation, 
which directly describes the dynamics of $\rho_{\rm M}(t)$ in the Majorana sector alone. 
To that end, we now trace also over the QD subspace.   
At long times, our numerical simulations generically show that $\rho(t)$
 factorizes into a Majorana part, $\rho_{\rm M}(t)$, and a QD contribution, $\rho_{\rm d}(t)$,
 \begin{equation}\label{factorize}
     \rho(t \to \infty) \simeq \rho_{\rm M}(t) \otimes \rho_{\rm d}(t).
 \end{equation}
The discussion in Sec.~\ref{sec2b} highlights that the
Majorana sector and the dot sector have to couple during intermediate times in order
to drive the Majorana system towards the desired target dark state (or dark space).
Once this state is stabilized, however, the dot electron can relax to the energetically favored state (up to the effects of the drive).
This argument also shows that, in accordance with our numerical observations, the specific choice for the tunneling parameters $\{\lambda_{j\nu}\}$ is only important 
for determining the targeted dark state while the 
disentanglement of Majorana and dot subspaces in Eq.~\eqref{factorize} represents a generic long-time feature.

For tracing over the QD part, we can effectively use a time-independent \emph{Ansatz},
\begin{equation}\label{ssform}
    \rho_{\rm d}=\left( \begin{array}{cc} 1-p & p_\perp \\ p^\ast_\perp & p  \end{array}\right), 
\end{equation}
written in the basis $\{ |0\rangle_{\rm d},|1\rangle_{\rm d}\}$ selected by the coupling to the QDs. Here, $p\ne 0$ refers to the  occupation probability of the energetically higher QD 2. This probability can be determined by numerically solving Eq.~\eqref{GeneralLindblad}, cf.~Sec.~\ref{sec2b}, or it may be treated as  phenomenological parameter.
A simple estimate predicts $p\approx {\rm max}(A,g_0)/\omega_0$.
Noting that a small but finite expectation value $\langle\tau_x\rangle\ne 0$ is observed in Fig.~\ref{fig2} at long times, we 
have also included an off-diagonal term $(p_\perp)$ in Eq.~\eqref{ssform}.

Inserting Eq.~\eqref{factorize} into Eq.~\eqref{GeneralLindblad} and tracing over the QD subsystem, we arrive at a 
Lindblad equation for the $2\times 2$ density matrix $\rho_{\rm M}(t)$ only,
\begin{equation}\label{LindbladMBQ}
\partial_t\rho_{\rm M}(t) = -i[\tilde H_{\rm L},\rho_{\rm M} ]+  \sum_{s=\pm}  \tilde\Gamma_s \mathcal{L}[\tilde J_s]\rho_{\rm M}(t),
\end{equation}
where the jump operators $\tilde J_\pm$ have been defined in Eq.~\eqref{jumpN}.
The dissipative transition rates $\tilde \Gamma_\pm$ in Eq.~\eqref{LindbladMBQ}  are given by
\begin{equation}\label{mod1}
  \tilde \Gamma_+ = p \Gamma_+,\quad \tilde \Gamma_- = (1-p)\Gamma_-,
\end{equation}
cf.~Eqs.~\eqref{dissrate} and \eqref{rate2}. 
The coherent time evolution in Eq.~\eqref{LindbladMBQ} is governed by the Hamiltonian 
\begin{equation}\label{mod22}
   \tilde H_{\rm L} =  2(1-2p)g_0 \tilde J_z + \sum_\pm \tilde h_\pm \tilde J_\pm^\dagger \tilde J_\pm^{},
\end{equation}
where $\tilde J_z$ has been specified in Eq.~\eqref{jzdef} and the Lamb shifts $\tilde h_\pm$ are given by 
\begin{equation}\label{mod2}
    \tilde h_+ = p h_+ ,\quad \tilde h_- = (1-p) h_-.
\end{equation}
The drive amplitude $A$ then appears only implicitly through the dependence $p=p(A)$.
We note that within the RWA, no contributions $\propto p_\perp$ appear in Eq.~\eqref{LindbladMBQ}.  Indeed, 
the RWA allows one to neglect terms $\propto \tau_+\rho\tau_+$ 
which stem from  $p_\perp\ne 0$.  

Importantly, apart from the initial transient behavior, 
all of our numerical results for the Majorana dynamics obtained from the full Lindblad equation for the combined QD-MBS system, Eq.~\eqref{GeneralLindblad}, are
quantitatively reproduced by using the simpler Lindblad equation \eqref{LindbladMBQ}. This statement is valid
for arbitrary model parameters subject to Eqs.~\eqref{condit} and \eqref{basiccond}.
We emphasize that the integration over the QD degrees of freedom as carried out above relies on the facts that (i) the convergence towards the 
target state is dictated by the Majorana sector, and that (ii) the QD and MBS degrees of freedom always decouple in the long-time limit, see Eq.~\eqref{factorize}. 
The latter feature has been established by extensive numerical simulations of Eq.~\eqref{GeneralLindblad}.
The reduced Lindblad equation \eqref{LindbladMBQ} is applicable
as long as transient behaviors are not of interest.  In particular,
when studying, e.g., the dynamics of $\rho_{\rm M}(t)$ in the presence of time-dependent QD level
energies $\epsilon_j(t)$, Eq.~\eqref{LindbladMBQ} should only be used for very slow (adiabatic) time dependences.  For rapidly varying QD level energies,
one has go back to the full Lindblad equation for the combined QD-MBS system in Eq.~\eqref{GeneralLindblad}.

\section{Dark state stabilization }\label{sec3}

Using the Lindblad master equation \eqref{LindbladMBQ} and the Choi isomorphism  \cite{Albert2014} summarized 
in App.~\ref{appB}, we now turn to a detailed analysis of our stabilization protocols for the single-box device in Fig.~\ref{fig1}. 
The parameter values  for stabilizing a specific dark state can  be determined by solving the zero-eigenvalue condition of the Lindbladian, cf.~App.~\ref{appB}. 
We recall that the key state design parameters of our DD protocol
 are given by the complex-valued tunneling amplitude parameters $\lambda_{j\nu}$, which also define the phases $\beta_j$ in Fig.~\ref{fig1}.
In Sec.~\ref{sec3a}, we show how to stabilize Pauli operator eigenstates.  In Sec.~\ref{sec3b}, we discuss magic state stabilization protocols, followed by a study of
 temperature effects in Sec.~\ref{sec3c}. 
 We show in Sec.~\ref{sec3d} that in certain cases, a dark state can be stabilized even in the absence of any driving field.
 Finally, we conclude in Sec.~\ref{sec3e} with several remarks.

\subsection{Pauli operator eigenstates}\label{sec3a}

 \begin{figure}[t]
\begin{centering}
\includegraphics[width=\columnwidth]{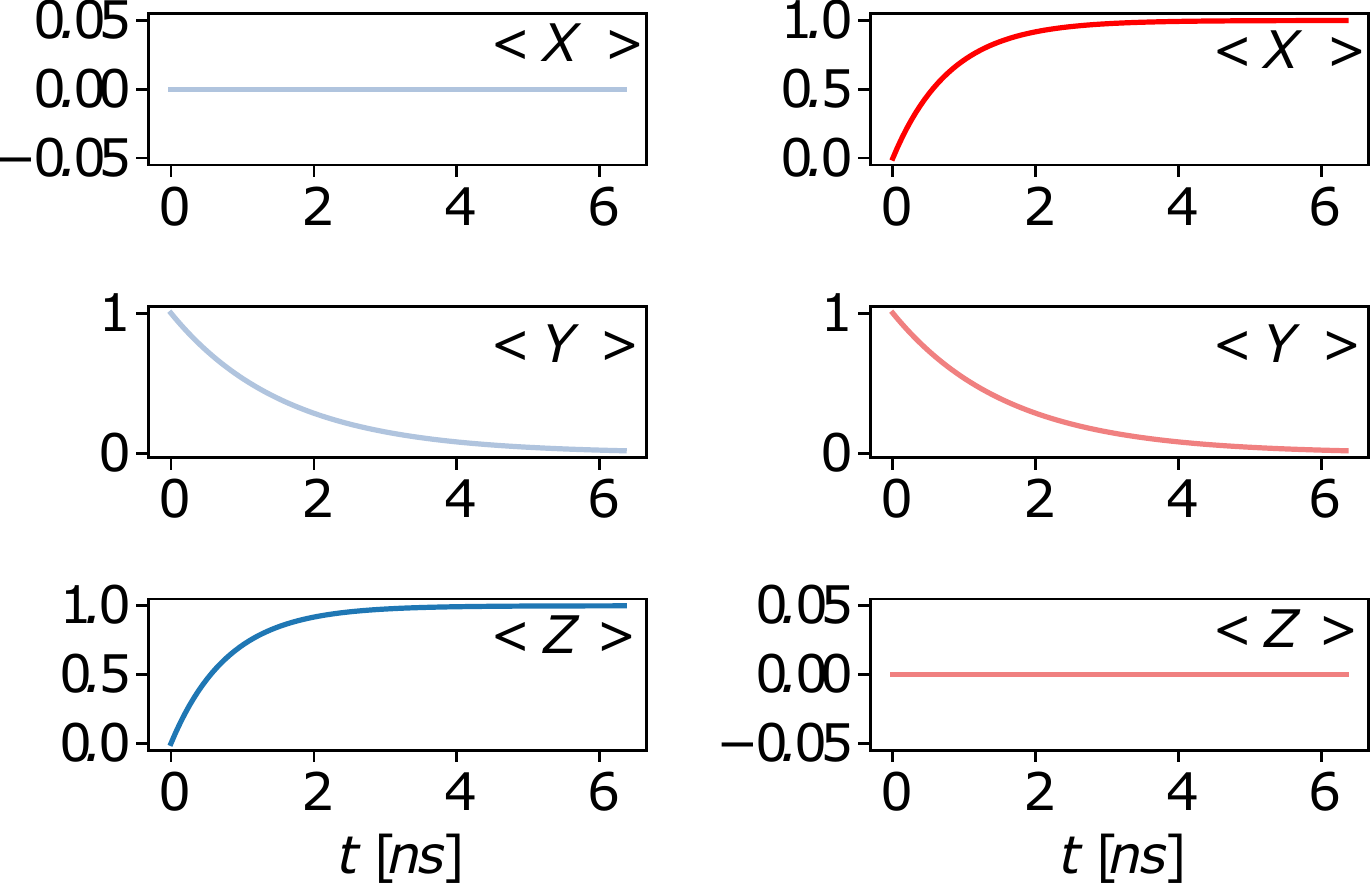}
\end{centering}
\caption{Dark-state stabilization protocols for Pauli operator eigenstates. Left side panels (blue curves): Stabilization of $|0\rangle$. Right side panels (red curves): Stabilization of $|+\rangle$, where $X|+\rangle=|+\rangle$.
In both cases, the Majorana island has initially been prepared in the $Y$-eigenstate with eigenvalue $+1$.  
We use the parameters in Eq.~\eqref{sigmazcond} with $p=1/2$, all other parameters are as in Fig.~\ref{fig2}. 
With $E_C=1$~meV and $g_0/E_C=2.5\times 10^{-3}$, the time units follow as shown.   As explained in the main text, for the chosen parameter set, Rabi oscillations are absent.
}
\label{fig4}
\end{figure}
 
We start by discussing DD protocols targeting Pauli operator eigenstates. Typical numerical results obtained by solving Eq.~\eqref{LindbladMBQ}
are illustrated in Fig.~\ref{fig4}.  Following the method in App.~\ref{appB}, the $Z=\pm 1$ eigenstates can be realized by choosing
\begin{equation}\label{sigmazcond}
|\lambda_{11}|=|\lambda_{12}|,\quad \lambda_{21}=\lambda_{22}=0, \quad \beta_1-\beta_3=\pm\pi/2, 
\end{equation}
with arbitrary  $\lambda_{23}$ and $\beta_{2,4}$, see Eq.~\eqref{betadef}. (We note that for $\lambda_{23}=0$, the phases $\beta_{2,4}$ are not defined.)  
At this point, we use the concept of a \emph{dissipative map} $\hat E$ \cite{Breuer2006}, which is defined in terms of a jump operator mapping the system onto a specific state when acting inside 
the Lindblad dissipator. For example, the dissipative maps targeting the $Z=\pm 1$ eigenstates are 
\begin{equation}
     \hat E_{\pm}=\sigma_\pm=(X\pm iY)/2. 
\end{equation}
For the stabilization parameters in Eq.~\eqref{sigmazcond},  the jump operator $\tilde J_+\propto \hat E_\pm$, with the $\pm$ sign determined by Eq.~\eqref{sigmazcond},
completely dominates the Lindbladian part of Eq.~\eqref{LindbladMBQ} at low temperatures, $T\ll\omega_0$.
 The dissipative dynamics then 
 maps every input state to $|0\rangle$ (for the $+$ sign) or $|1\rangle$ (for the $-$ sign).  At the same time, the Hamiltonian evolution in Eq.~\eqref{LindbladMBQ} comes from $\tilde H_{\rm L}\propto Z$, see Eq.~\eqref{mod22}. Evidently, this  Hamiltonian commutes with the targeted state $\rho_{\rm M}(\infty)$, and therefore does not affect the dynamics towards the steady state generated by the dissipative map $\hat E_\pm$.
 The Majorana state $\rho_{\rm M}(t)$ is thus automatically 
steered towards the corresponding $Z$-eigenstate by the Lindbladian, with no obstruction from the Hamiltonian  dynamics.  

For the above protocol, the \emph{dissipative gap} is given by, cf.~App.~\ref{appB}, 
\begin{equation}\label{deltaz}
    \Delta_z = |4\lambda_{11}\lambda_{23}|^2\sum_{s=\pm}\tilde\Gamma_{s}.
\end{equation}
In general terms, the dissipative gap is defined as the real part of the smallest non-vanishing eigenvalue of the Lindbladian (the dark state itself has eigenvalue zero) \cite{Breuer2006}. 
 The time scale on which the dark state will be approached is therefore given by $\Delta_z^{-1}$.  
 Moreover, the approach of the Bloch vector towards the dark state $|0\rangle$ is in general accompanied by damped oscillations in the $(X,Y)$ components, where $\Delta_z$ is the damping rate and the Rabi frequency follows from Eq.~\eqref{mod22}  as
 \begin{equation}\label{rabiz}
 \Omega_z \simeq \left|2g_0(1-2p)|\lambda_{11}|^2-8|\lambda_{11}\lambda_{23}|^2\tilde h_+\right|.
 \end{equation}
 For the special case $\lambda_{21}=0$ with $p=1/2$, cf. Sec.~\ref{sec2b}, and noting that $\tilde h_+=0$ for $\alpha=1/4$, cf.~Eq.~\eqref{rate2},
 we obtain $\Omega_z=0$ in Eq.~\eqref{rabiz}. The left panels in Fig.~\ref{fig4} therefore exhibit only damping in the $(X,Y)$ components, without Rabi oscillations. 

 Next, $X=\pm 1$ eigenstates are realized by choosing
\begin{equation}
|\lambda_{21}|=|\lambda_{23}|,\quad \lambda_{11}=\lambda_{22}=0,\quad \beta_2=\mp\pi/2,
\end{equation}
with the dissipative gap $\Delta_x=|4\lambda_{12}\lambda_{21}|^2\sum_s\tilde\Gamma_s.$
As shown in the right panels of Fig.~\ref{fig4},  $X$-eigenstates, e.g., the state $|+\rangle$ 
for eigenvalue $+1$, 
can be stabilized using the setup in Fig.~\ref{fig1}.   As for the $Z$-stabilization shown in the left panels, there are no Rabi oscillations for this parameter set.  

Finally, for stabilizing the $Y$-eigenstates with eigenvalue $\pm 1$, one requires
\begin{equation}
|\lambda_{22}|=|\lambda_{23}|, \quad \lambda_{12}= \lambda_{21}=0,\quad \beta_2-\beta_3-\beta_4=\pm\pi/2, 
\end{equation}
with the dissipative gap $\Delta_y = |4\lambda_{11}\lambda_{22}|^2\sum_s\tilde\Gamma_s$. 

In all these examples, the target axis (say, $\hat e_z$ for $Z$-eigenstates) is controlled by selecting appropriate tunneling amplitude parameters $\lambda_{j\nu}$.  Two links are switched off, and two are matched in amplitude such that the desired jump operator $\tilde J_+$ is implemented.
For $T\ll \omega_0$, dissipative transitions are fully governed by this jump operator which is due to inelastic cotunneling transitions from QD $2\to 1$.  
Under these conditions, we find that $\tilde H_{\rm L}$ commutes with the Pauli operator $\hat\sigma$ corresponding to the target axis (e.g., $\hat\sigma=Z$ for $Z$-states).   
Finally, by adjusting the phases $\beta_j$, one can select the stabilized state, say, $|0\rangle$ or $|1\rangle$. 
It is a remarkable feature of our Majorana-based DD setup that the Hamiltonian $\tilde H_{\rm L}$ can be engineered to only generate $\hat\sigma$.  As a consequence, the Lindbladian dissipator already drives the system to the desired dark state.  

\subsection{Magic states} \label{sec3b}

\begin{figure}[t]
\begin{centering}
\includegraphics[width=\columnwidth]{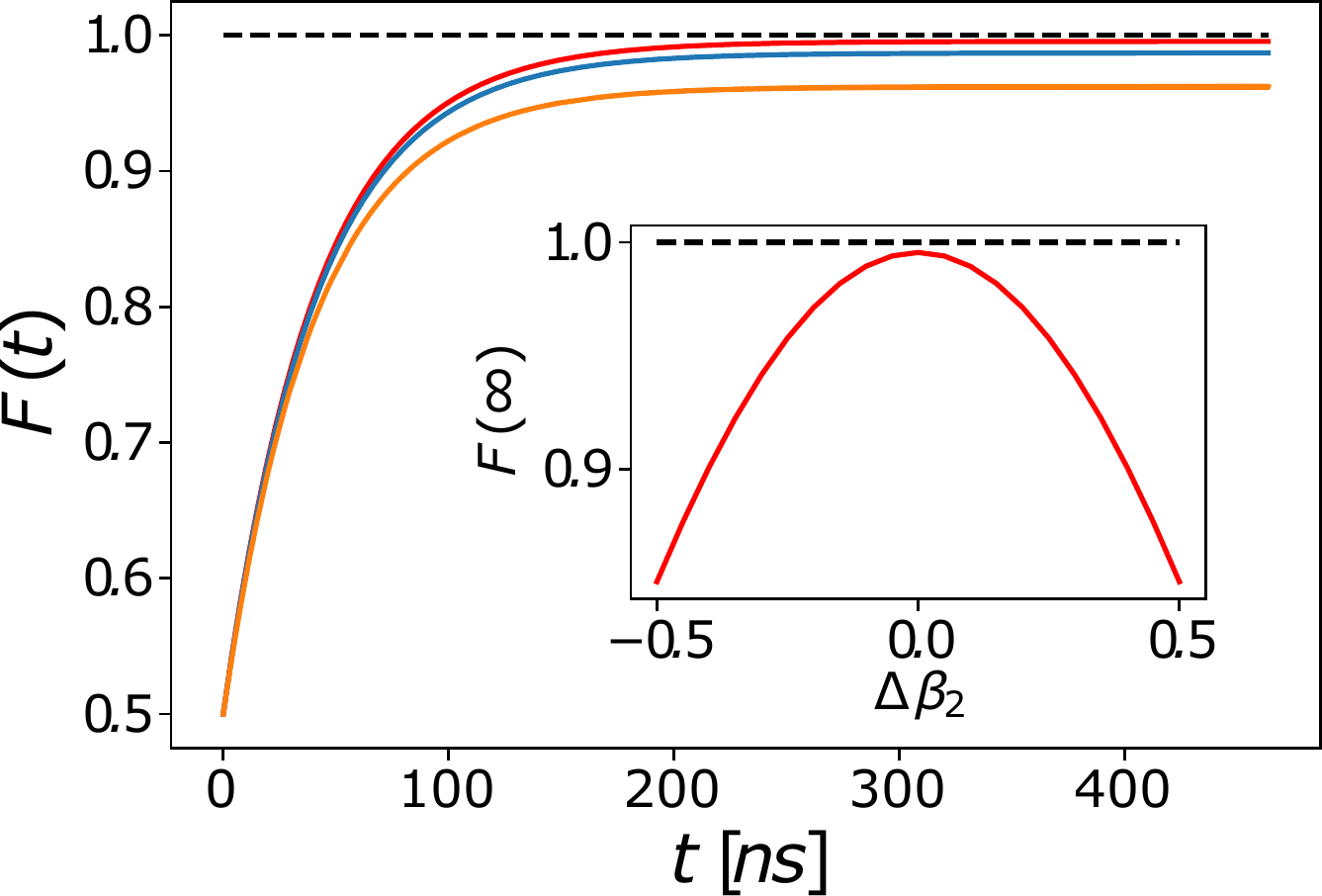}
\end{centering}
\caption{ Fidelity for a stabilization protocol targeting the magic state $|m\rangle$. Here
 the Majorana state follows by numerical integration of Eq.~\eqref{LindbladMBQ} using the parameters 
in Eq.~\eqref{magiccond} with $|\lambda_{23}|=1$. 
Other parameters are $E_C = 1$~meV, $g_0/E_C=2.5\times 10^{-3}, T/g_0=4$, 
$\omega_0/g_0=40$, $\omega_c/g_0 =200$, $\alpha=1/4$, and $p = 0.01$. 
Main panel: Time dependence of the fidelity for ideal parameters [Eq.~\eqref{magiccond}] (red curve), with  a mismatch of order $10\%$ in all state design parameters [$|\lambda_{11}|=-0.1+1/\sqrt2,|\lambda_{21}|=+0.1+1/\sqrt{2},|\lambda_{12}|=|\lambda_{22}|=0.9,\beta_3=-\beta_2=11 \pi/20$] (blue), and a mismatch of order $20\%$ in the same  parameters (orange).
Inset: Steady-state fidelity vs deviation $\Delta\beta_2$ with otherwise ideal parameters, where $\beta_2=-\frac{\pi}{2}(1+\Delta\beta_2)$.
\label{fig5}}
\end{figure}

In order to highlight the power of our DD stabilization protocols, we next consider the magic state \cite{Nielsen}
\begin{equation}\label{magicstate}
    |m\rangle = e^{-i\frac{\pi}{8}Y} |0\rangle.
\end{equation}
The practical importance of this state comes from the fact that a large number of ancilla qubits approximately prepared in the state $|m\rangle$ 
are needed for the magic state distillation protocol.  The latter is an essential ingredient for implementing the $T$-gate required for universal surface code quantum computation
\cite{Fowler2012,Vijay2015,Landau2016,Plugge2016,Nielsen}. 
Targeting $|m\rangle$, the stabilization conditions now involve all tunnel links in Fig.~\ref{fig1} and are given by
\begin{eqnarray}  \label{magiccond}
    |\lambda_{12}|&=&|\lambda_{23}|, \quad |\lambda_{21}|=|\lambda_{11}|=|\lambda_{23}|/\sqrt2,\\ \nonumber \lambda_{22}&=&0,\quad \beta_3=\beta_1+\beta_2, \quad \beta_2 = -\pi/2.
\end{eqnarray}
We here define the \emph{fidelity} of the state $\rho_{\rm M}(t)$ with respect to a specific pure state, $\rho_{\rm M}^{(0)}=|\psi\rangle\langle \psi|$, as
\begin{equation} \label{fidelity}
    F(t)={\rm tr}\left[ |\psi\rangle\langle \psi|\rho_{\rm M}(t)\right].
\end{equation}
We show numerical results for the magic state fidelity with $|\psi\rangle=|m\rangle$ in Fig.~\ref{fig5}, 
using the parameters in Eq.~\eqref{magiccond}.
We find $F=1$ at long times for the ideal parameter choice in Eq.~\eqref{magiccond}.  Figure \ref{fig5} also illustrates the long-time fidelity when allowing for
small deviations from Eq.~\eqref{magiccond} which are inevitable in practical implementations.  Remarkably, even for sizeable deviations from the ideal parameter set, the fidelity remains close to unity.   By determining the spectrum of the Lindbladian, we obtain the dissipative gap as
\begin{equation}\label{dissgapm}
    \Delta_m = |4\lambda_{11}\lambda_{23}|^2\sum_s\tilde \Gamma_s.
\end{equation}
 Using the parameters in Fig.~\ref{fig5}, we find $\Delta_m^{-1}\simeq 80$~ns.
Even though our magic state stabilization protocol requires more parameter fine tuning than the stabilization of $|0\rangle$, 
the dark state $|m\rangle$ is reached on  essentially the same time scale.

\subsection{Effect of temperature}\label{sec3c}

\begin{figure}[t]
\begin{centering}
\includegraphics[width=0.95\columnwidth]{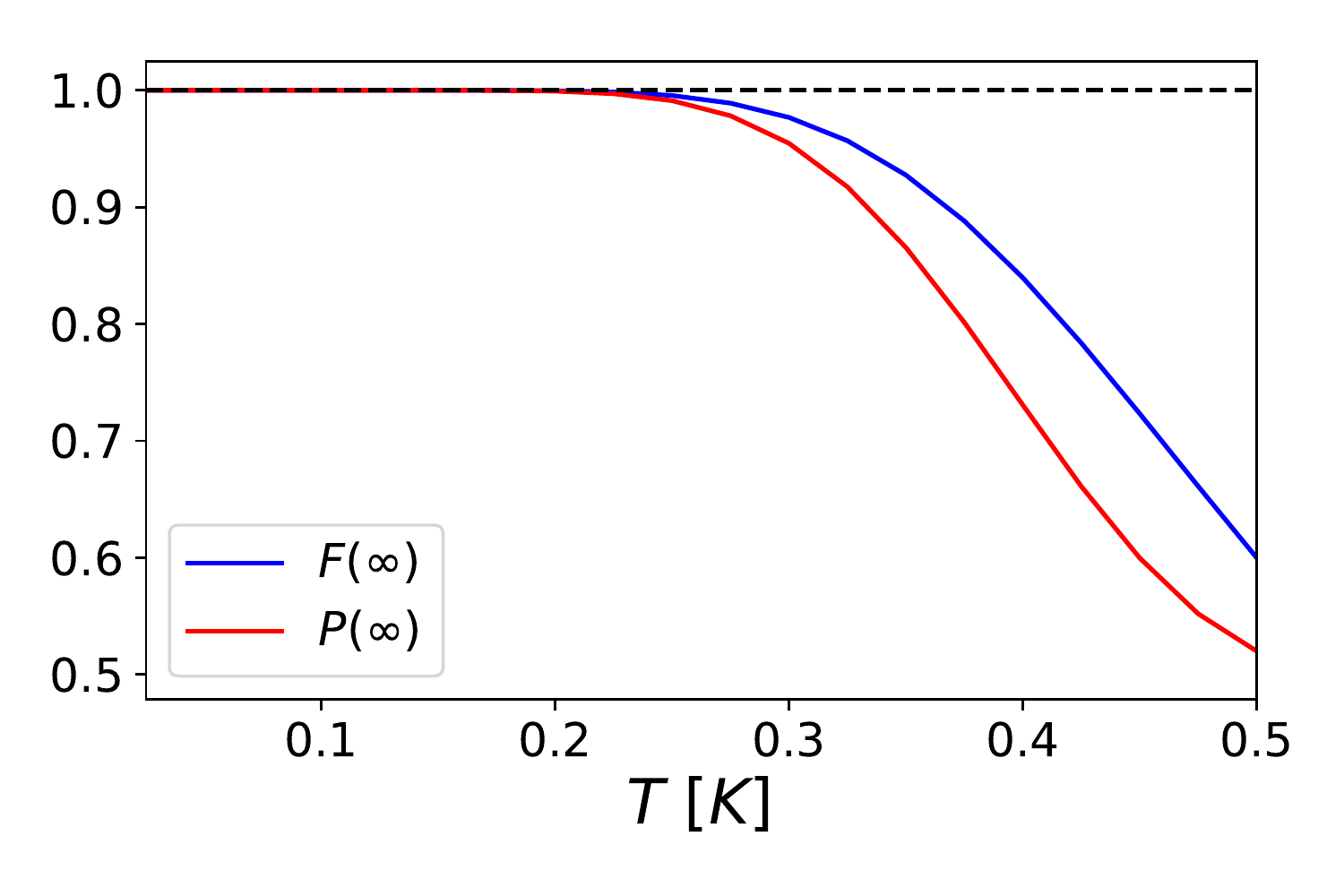}
\end{centering}
\caption{ Steady-state fidelity, $F(\infty)$, and purity, $P(\infty)$, vs temperature (in Kelvin) for 
the state $|0\rangle$ and for the magic state $|m\rangle$. We use ideal state design parameters, see Eqs.~\eqref{sigmazcond} and \eqref{magiccond}, with all other parameters as in 
Figs.~\ref{fig4} and \ref{fig5}, respectively. The numerical results for both states cannot be distinguished for these parameter choices on 
the shown scales. The frequency $\omega_0$ corresponds to a temperature of $\approx 2.5$~K, while $E_C=1$~meV \cite{Lutchyn2018} corresponds to $\approx 11$~K. }
\label{fig6}
\end{figure}

We next address the effect of raising temperature within the conditions set by Eq.~\eqref{basiccond}, in particular $T\ll \omega_0$.  
Figure~\ref{fig6} shows numerical results for the $T$-dependent steady state
fidelity $F(\infty)$ with respect to the states $|0\rangle$ and  $|m\rangle$,
choosing ideal parameters as in Eqs.~\eqref{sigmazcond} and  \eqref{magiccond}, respectively.

At very low temperatures,
the fidelity stays very close to the ideal value ($F=1$) since here only the rate
 $\tilde \Gamma_+$, see Eqs.~\eqref{dissrate} and \eqref{mod1}, is significant. In this limit, corrections 
to $F=1$ are exponentially small and appear to be governed by the dissipative gap, $1-F\propto \exp(-\Delta_{z/m}/T)$. The same scaling behavior also applies to the purity.
As temperature increases, the thermal excitation
rate $\tilde\Gamma_-=e^{-\omega_0/T}\tilde \Gamma_+$  cannot be neglected anymore. Focusing on the stabilization of the state $|0\rangle$, we have $\tilde J_-\propto \sigma_-$.  The Lindblad dissipator $\tilde \Gamma_- {\cal L}[\tilde J_-]$ will then target the `wrong' $Z$-eigenstate $|1\rangle$. The competition between ${\cal L}[\tilde J_+]$ and ${\cal L}[\tilde J_-]$ implies that the fidelity will deteriorate as temperature increases.

This expectation is confirmed by our numerical results.
For the parameters in Fig.~\ref{fig6}, the fidelity noticeably drops once 
$T$ exceeds the crossover temperature $T_c\approx 250$~mK.
Figure \ref{fig6} also shows the temperature dependent purity of the steady state, 
$P(\infty) ={\rm tr}\rho_{\rm M}^2(t\to \infty)$.  For $T\ll T_c$,
we find $P(\infty) \simeq 1$.   As $T$ increases, however, the maximally mixed state $\rho_{\rm M}(\infty)=\frac12 \mathbb{1}$ with $F(\infty)= P(\infty)=1/2$ is approached, and consequently the purity also becomes smaller.

Finally, let us note that at elevated temperatures, the RWA will also become less accurate. One may thus need to account for dephasing effects induced by corrections beyond RWA \cite{Shavit2019}. However, for the results shown in Fig.~\ref{fig6} with $T/\omega_0<0.2$, such effects are expected to be very small.

\subsection{Stabilization without driving field} \label{sec3d}

In certain cases, it is possible to stabilize dark states even without drive Hamiltonian, $H_{\rm drive}=0$.
In this subsection, we demonstrate the feasibility of this idea for special choices of the state design parameters.
We are not aware of other DD systems allowing for dark states in the absence of driving.  
In our setup, we will see that the dissipative dynamics can also generate terms that mimic the effects of a weak driving field.

To be specific, we apply the Lindblad equation \eqref{LindbladMBQ} to 
setups where $\lambda_{j\nu}\ne 0$ only for $(j \nu) \in \{ 11,12,23 \}$.
In particular, since $\lambda_{21}=0$, this case corresponds to the special parameter regime discussed in Sec.~\ref{specialsec}.
For simplicity, below we  drop the exponentially small contribution to the dissipator due to $\tilde J_-$.
From Eq.~\eqref{jumpN}, the only relevant jump operator is then given by
\begin{equation} \label{jplus1}
\tilde J_{+} = 2i  \lambda_{23}^\ast \left( \lambda_{11} X - \lambda_{12} Y \right).
\end{equation}
In addition, we keep Lamb shift effects implicit. In particular,
 they can be taken into account by renormalizing $B_z$ in Eq.~\eqref{HM1} below.
The operator $\tilde J_z$ entering $\tilde H_{\rm L}$, see Eqs.~\eqref{jzdef} and \eqref{mod22}, 
has the form
\begin{equation}\label{jz1}
\tilde J_z  =-\sin\beta_1\left| \lambda_{11} \lambda_{12} \right|  \, Z .
\end{equation}
We now study the undriven ($A=0$) scenario for two parameter sets allowing for analytical progress.
The stabilization of pure dark states may then be possible because the Hamiltonian $\tilde H_L$ can 
effectively take over the role of the drive. As a result, the arguments behind the factorized form of the long-time 
density matrix in Eq.~\eqref{factorize} carry over to the present case.
The frequency $\omega_0$ now simply represents the (positive) energy difference $\epsilon_2-\epsilon_1$, see Eq.~\eqref{omegadef}, instead of a drive frequency.
Moreover, we assume $p_\perp = 0$  while the probability $p$ in Eq.~\eqref{ssform} is estimated by $p\approx g_0/\omega_0$. 
We note in passing a finite static contribution to the inter-QD tunnel coupling, $t_{12}\ne 0$ in Eq.~\eqref{Hdriv}, can be taken
into account here.  This coupling will modify $p$ according to $p \approx{\rm max}(t_{12}, g_0)/\omega_0$.
We also recall that for $A\ne 0$, one instead finds $p=1/2$ since we have $\lambda_{21}=0$, cf.~Sec.~\ref{specialsec}.

\subsubsection*{Case 1: $\lambda_{11}=\pm i \lambda_{12}$}

The first case is defined by $\lambda_{11}=is \lambda_{12}$, with $s=\pm 1$.
We observe that the dot fermion operator $d_1$ corresponding to QD 1 is then tunnel-coupled to a nonlocal fermion formed from the Majorana operators, 
$c=(\gamma_1-is \gamma_2)/2$.  With $\sigma_\pm=(X\pm iY)/2$, Eqs.~\eqref{jplus1} and \eqref{jz1} simplify to 
\begin{equation}\label{case1}
\tilde J_+=4i\lambda_{23}^\ast \lambda_{11}\sigma_{-s}, \quad\tilde J_z=-s |\lambda_{11}|^2 \, Z.
\end{equation}
The Lindblad equation \eqref{LindbladMBQ} is then given by
\begin{equation}\label{nodrive:LBMcase1}
\partial_t \rho_{\rm M}(t)  = - i [ \tilde H_{\rm L}, \rho_{\rm M}(t) ] +
\Gamma_1  {\cal L} \left[ \sigma_{-s} \right] \rho_{\rm M}(t),
\end{equation}
where the Hamiltonian follows from Eq.~\eqref{mod22} as
\begin{equation}\label{HM1}
\tilde H_{\rm L} = -2s (1-2p) g_0 |\lambda_{11}|^2 Z=s B_z Z.
\end{equation}
We note that the Lamb shift $\tilde h_+$ can be taken into account by redefining $B_z$.
Furthermore, the rate $\Gamma_1$ in Eq.~\eqref{nodrive:LBMcase1} is proportional to $\tilde \Gamma_+$ in Eq.~\eqref{rate2}.
The only zero eigenstate of the Lindbladian is the $Z$-eigenstate $|0\rangle$ (for $s=-1$) [or $|1\rangle$ (for $s=+1$)], e.g., 
${\cal L} \left[ \sigma_+ \right] |0 \rangle \langle 0 | = 0$.
The same $Z$-eigenstate is also the lowest energy eigenstate of $\tilde H_{\rm L}$ in Eq.~\eqref{HM1}.

Using the $Z$-eigenstate basis $\{ |0\rangle , |1\rangle \}$ for $s=-1$ [and $\{|1\rangle,|0\rangle \}$ for $s=+1$], we 
can parametrize the time-dependent density matrix $\rho_{\rm M}(t)$ solving Eq.~\eqref{nodrive:LBMcase1} with real-valued $x(t)$ subject to $0\le x\le 1$ and 
complex-valued $y(t)$ as
\begin{equation}
\rho_{\rm M}(t)  = \left( \begin{array}{cc}  1 - x(t) & y(t) \\ y^\ast(t) &x(t)\end{array}\right).
\end{equation}
The quantities $x(t)$ and $y(t)$ represent the diagonal and off-diagonal density matrix deviations, respectively, from the steady-state density matrix corresponding to the stabilized 
$Z$-eigenstate. Using Eq.~\eqref{nodrive:LBMcase1}, these deviations obey the equations of motion
\begin{equation}\label{relax1}
\partial_t x = - \Gamma_1 x,\quad \partial_t y = - 2 i B_z y - \frac{\Gamma_1}{2} y,
\end{equation}
which explicitly shows the relaxation and decoherence dynamics of $\rho_{\rm M}(t)$ towards the stabilized pure
state.  The above example demonstrates that the dissipative stabilization of a dark state can be achieved even in the absence of a driving field  in our Majorana box setup.

\subsubsection*{Case 2: $\beta_1=0$}

Putting the phase $\beta_1$ to zero, $d_1$ is effectively coupled to a single Majorana operator, $\gamma_{\rm eff}=\gamma_1 \cos \delta + \gamma_2\sin \delta$, 
with $\delta= \tan^{-1} \left| \lambda_{12} / \lambda_{11} \right|$.
One then obtains $\tilde J_z=0$.  The jump operator 
$\tilde J_+$ is now given by
\begin{equation}\label{case2}
 \tilde J_+= B_{\perp}\sigma_+ e^{i\delta}+{\rm h.c.},\quad 
B_\perp= 2i \lambda_{23}^\ast \lambda_{11}/|\cos\delta|.
\end{equation}
Noting that the Lamb shifts in $\tilde H_{\rm L}$ only give an irrelevant constant, we arrive at the Lindblad equation 
\begin{equation}\label{nodrive:LBMcase2}
\partial_t \rho_{\rm M}(t)  = \frac{\Gamma_2}{4} {\cal L} \left[ \sigma_{\bf n} \right] \rho_{\rm M}(t),
\end{equation}
where we define
\begin{equation}
\sigma_{\bf n} = {\bf n \,\cdot} {\bm \sigma} = \sigma_+ e^{i \delta} + \sigma_-e^{-i\delta},
\end{equation}
with the unit vector ${\bf n} = (\cos \delta, - \sin \delta, 0)$. Again, the rate $\Gamma_2$ is proportional
to the respective rate $\tilde \Gamma_+$ in Eq.~\eqref{rate2}.

For the case in Eq.~\eqref{nodrive:LBMcase2}, the Lindbladian has two zero eigenstates,
${\cal L} \left[ \sigma_{\bf n} \right] | s \rangle \langle s | =
{\cal L} \left[ \sigma_{\bf n} \right] | a \rangle \langle a | = 0$, corresponding to the 
eigenstates of ${\bf \sigma}_{\bf n}$, i.e.,
$\sigma_{\bf n}| s \rangle = | s \rangle$ and $\sigma_{\bf n}| a \rangle =- | a \rangle$.
Using the $X$-eigenstates $|\pm\rangle$, one finds
\begin{equation}
| s / a \rangle =\frac{1}{ \sqrt{2}} \left( e^{i \delta} |+\rangle \pm e^{-i \delta} |-\rangle \right).
\end{equation}
In the $\{ |s\rangle,|a\rangle\}$ basis, $\rho_{\rm M}(t)$ can  be parametrized as 
\begin{equation}
\rho_{\rm M}(t)  =  \left( \begin{array}{cc} \frac12 + x(t) & y(t) \\ y^\ast(t) & \frac12-x(t)
\end{array} \right),
\end{equation}
where the real-valued parameter $x(t)$ has to satisfy $|x|\le 1/2$. Equation~\eqref{nodrive:LBMcase2} then yields 
\begin{equation}
\partial_t x = 0,\quad \partial_t y = - \frac{\Gamma_2}{2} y.
\end{equation}
Clearly, there is no relaxation in the basis selected by the environment via the QDs, i.e.,  $x(t)$
remains constant.  Only the off-diagonal elements of the density matrix are subject to decay with the rate $\Gamma_2/2.$  One can therefore prepare an arbitrary mixed state as steady state.

\subsection{Discussion}\label{sec3e}
 
We conclude this section with several additional points.

\subsubsection{Mixed states}

As pointed out in Sec.~\ref{sec3d}, one can also use our protocols for stabilizing mixed states, see also Ref.~\cite{Kumar2020}.   
To give another example, now for $A\ne 0$, we consider changing the above phase conditions such that a mixture of Pauli eigenstates can be prepared as dark state.  For instance, by choosing the state design parameters as in Eq.~\eqref{sigmazcond} but keeping $\bar\beta=\beta_1-\beta_3$ arbitrary, one obtains the dark state
 \begin{equation}
    \rho_{\rm M}(\infty)=\frac{1+\sin\bar\beta}{2}|0\rangle\langle 0|+\frac{1-\sin\bar\beta}{2}|1\rangle\langle 1|.
\end{equation}
The relative weight of the two components can then be altered by adjusting the phase difference $\bar\beta$.  

\subsubsection{Majorana overlaps}

So far we have assumed that the overlap between different MBSs is negligibly small.
What are the effects of a finite (but small) hybridization between different MBS pairs on the above stabilization protocols?  
Such terms could arise, e.g., due to the finite nanowire length \cite{Alicea2012}.
They are described by a Hamiltonian term $ H'=\sum_{\nu<\nu'}i\epsilon_{\nu\nu'}\gamma_\nu\gamma_{\nu'}$, with
hybridization energies $\epsilon_{\nu\nu'}$.
By construction, such a term survives the RWA and the Schrieffer-Wolff projection in Sec.~\ref{sec2} and thus contributes to the Hamiltonian $\tilde H_{\rm L}$ in the Lindblad equation \eqref{LindbladMBQ} without affecting the Lindbladian dissipator. 
In the Pauli operator language, such terms act like a weak magnetic Zeeman field.  
If the corresponding field is parallel to the target axis of the dark state, 
it does not cause any dephasing.  For instance, for the stabilization of the $Z$-eigenstate $|0\rangle$, 
the hybridization parameters $\epsilon_{12}$ and $\epsilon_{34}$ can be tolerated as they only couple to the Pauli operator $Z$ in Eq.~\eqref{PauliOp}.
Clearly, such couplings have no detrimental effects on our stabilization protocols. 
For the stabilization of arbitrary target states, however, the role of MBS overlaps is more subtle, in particular when power-law scaling of the overlap with increasing distance becomes important \cite{Aseev2019}.  A detailed discussion of such effects will be given elsewhere.

\subsubsection{Readout dynamics}

For reading out a stabilized dark state, it is possible to use the same techniques suggested previously for the native Majorana qubit \cite{Plugge2017,Karzig2017,Munk2019}.
In particular, one can perform capacitance spectroscopy using
additional single-level QDs that are tunnel-coupled to MBS pairs. These QDs are used for measurements only, where
the spectroscopic signal contains an interference term which depends on the respective Pauli matrix in Eq.~\eqref{PauliOp}. 
This projective readout yields the Pauli eigenvalue $\pm 1$  with a state-dependent probability \cite{Karzig2017}. 
Of course, this method can also be used to prepare the Majorana island in  a Pauli eigenstate before the DD protocol is started.
In order for the readout not to interfere with the DD stabilization protocol, one has to make sure that the characteristic projective measurement time 
scale (see Refs.~\cite{Plugge2017,Karzig2017} for detailed expressions) is much longer than the typical
inelastic cotunneling time $\tilde\Gamma_+^{-1}$. 
Similarly, single-electron pumping protocols via a pair of QDs attached to different MBSs 
allow one to apply a Pauli operator to the tetron state in a topologically protected manner \cite{Plugge2017}.  

\subsubsection{Beyond the horizon of a dark state}
\label{sec3e4}

So far we have discussed DD stabilization protocols targeting a desired dark state. 
The dark space dimension for those protocols is $D=1$, see App.~\ref{appC}.
Since there is a unique dark state for a given choice of the state design parameters, 
one could utilize a DD single-box device as a self-correcting quantum memory.  
By means of adiabatic changes of the state design parameters, one can in principle steer the Majorana state on its Bloch sphere. However, for general 
state manipulation protocols, it is advantageous to have access to 
a dark space manifold with $D>1$, which may be engineered in systems with more than four MBSs. 
We address this case in the next section.

\section{Dark space engineering }
\label{sec4}

\begin{figure}[t]
\begin{centering}
\includegraphics[width=\columnwidth]{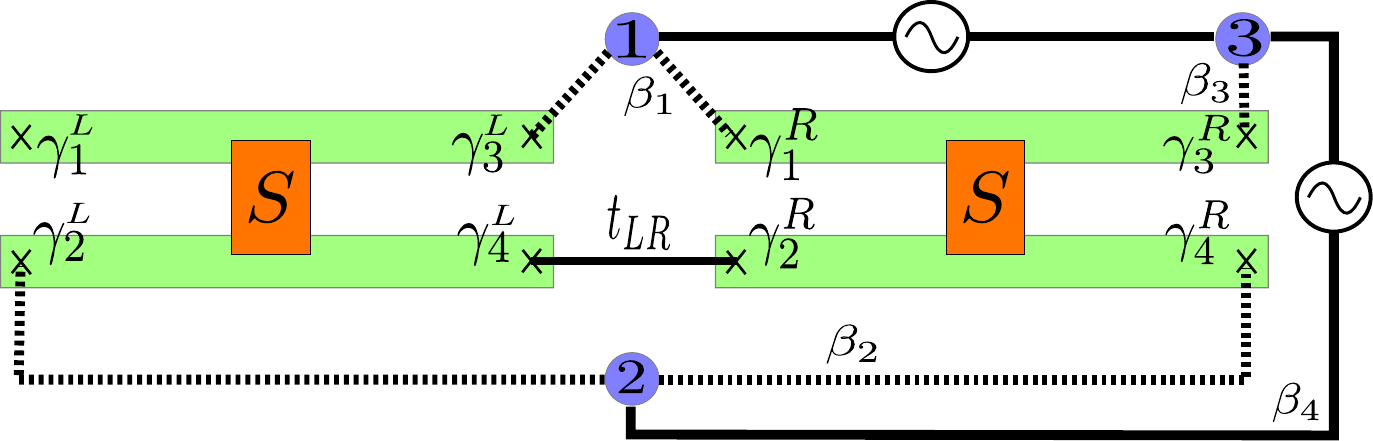}
\end{centering}
\caption{Schematic two-box layout for DD dark space stabilization and manipulation protocols, cp.~Fig.~\ref{fig1} for the single-box case. 
The left (right) box harbors four MBSs described by $\gamma_{\nu}^L$ ($\gamma^R_\nu$).  
The tunneling bridge with amplitude $t_{LR}$  connects $\gamma_4^L$ and $\gamma_2^R$.
QD 3 has independently driven tunneling bridges to QD 1 and to QD 2 (solid lines).  The three QDs are operated in
the single-electron regime, $N_{\rm d}=1$.
The electromagnetic environment affects the phases of the tunnel links betweens QDs and MBSs 
(dashed lines).  The phases $\beta_{j}$ for this geometry are also indicated.}
\label{fig7}
\end{figure}

We continue with DD  protocols targeting quantum states within a dark space manifold. A degenerate manifold of dark states may 
be engineered by employing a device with at least two Majorana boxes as depicted in Fig.~\ref{fig7}.
After introducing our model and the 
corresponding Lindblad equation in Sec.~\ref{sec4a}, we show in Sec.~\ref{sec4b} how a dark space can be created and classified. 
In Sec.~\ref{sec4c}, we then describe how to stabilize Bell states.  In Ref.~\cite{ourprl}, we describe external perturbations for moving the dark state to another state within the protected dark space manifold, and we show how to create a dark space manifold realizing a `dark Majorana qubit'.  In such a system, topological and DD mechanisms reinforce each other and thereby can provide exceptionally high levels of fault tolerance. 
Moreover, we remark that the stabilization of Bell states can also be implemented in a hexon device (i.e., a Majorana box with six MBSs \cite{Karzig2017}), see Ref.~\cite{GauThesis}. 

\subsection{Lindblad equation for two coupled boxes}\label{sec4a}

\subsubsection{Model}

Following the discussion in Sec.~\ref{sec2a}, we describe the two islands in Fig.~\ref{fig7} by  $H_{\rm box}=H_{{\rm box},L}+H_{{\rm box},R}$, with $H_{{\rm box},L/R}$ as in Eq.~\eqref{Hbox}. Here, the four MBSs on the left (right) box 
correspond to Majorana operators $\gamma_\nu^L$ ($\gamma_\nu^R$).  Both islands are separately operated under Coulomb valley conditions.
For notational simplicity, we assume that they have the same charging energy, $E_{C,L}=E_{C,R}=E_C$.     
Focusing on the long-wavelength components of the electromagnetic environment, we again work with a single bosonic bath, $H_{\rm env}=\sum_m E_m b_m^\dagger b_m^{}$, where photons couple to the QDs and MBSs via fluctuating phases, $\theta_j$, in the tunneling Hamiltonian, see Sec.~\ref{sec2a}.
The setup in Fig.~\ref{fig7} requires up to three single-level QDs, 
$H_{\rm d}=\sum_{j=1}^3 \epsilon_j d_j^\dagger d_j^{}$,
where QD 3 couples to both other QDs via independently driven tunnel links.   
We consider the regime $N_{\rm d}=1$, where on time scales $\delta t>1/E_C$, the three QDs share a single electron.

Using the interaction picture with respect to the dot Hamiltonian $H_{\rm d}$, the full Hamiltonian is then given by
\begin{equation}\label{Hfull2MBQ}
H(t) = H_{\rm box} + H_{\rm env}+ H_{LR} + H_{\rm drive}(t) + H_{\rm tun}(t),
\end{equation}
where a phase-coherent tunnel link couples the boxes. Without loss of generality, we assume a real-valued tunneling amplitude $t_{LR}>0$, 
\begin{equation}\label{LRCoup}
    H_{LR} = i t_{LR} \gamma_4^L \gamma_2^R.
\end{equation}
The drive Hamiltonian now has the form
\begin{equation}\label{Hdrive2}
    H_{\rm drive}(t)=\sum_{j=1,2} 2A_j\cos\left(\omega_jt\right)e^{i\left(\epsilon_j-\epsilon_3\right)}d_j^{\dagger}d^{}_3 + {\rm h.c.},
\end{equation}
where the two driving fields have the respective amplitude $A_{1,2}$ and frequency $\omega_{1,2}$.
In analogy to Eq.~\eqref{Htun}, the QD-MBS tunnel links are described by  
\begin{equation}\label{Htun2}
    H_{\rm tun}(t)=t_0 \sum_{j\nu, \kappa=L/R}\lambda_{j,\nu\kappa}^{} e^{-i\phi_\kappa} e^{i\theta_j} e^{i\epsilon_jt} d^\dagger_j  
    \gamma_{\nu}^\kappa  +{\rm h.c.},
\end{equation}
with the phase operators $\phi_{L/R}$ for the left/right Majorana island.  Using the same approximations as in Sec.~\ref{spsec1},
the electromagnetic environment enters Eq.~\eqref{Htun2} through the 
fluctuating phases $\theta_j$. 
 With the overall energy scale $t_0$, the complex-valued 
parameters $\lambda_{j,\nu\kappa}$ parametrize the transparency of the tunnel contact between
$d_j$ and  $\gamma_\nu^{\kappa=L/R}$. Similar to Eq.~\eqref{betadef}, the phases $\beta_j$ in Fig.~\ref{fig7}
follow from the phases of these parameters. Since $\beta_4$ can be absorbed by a renormalization of $\beta_3$ for the purposes below, we put $\beta_4=0$.

To simplify the presentation, we next assume that QDs 1 and 2 have the same 
energy level, $\epsilon_1=\epsilon_2$. Moreover, we consider the case of 
equal drive frequencies, $\omega_1=\omega_2\equiv \omega_0$, and identical drive amplitudes, $A_1=A_2\equiv A$,
and again impose a resonance condition, $\omega_0=\epsilon_3-\epsilon_1$. 
However, in analogy to our discussion in Sec.~\ref{sec2}, we expect that 
overly precise fine tuning with respect to those conditions is not necessary.  

We now proceed in analogy to Sec.~\ref{sec2a} with the construction of an effective low-energy model by means of a Schrieffer-Wolff transformation
to the lowest-energy charge state in each box.  We can then define Pauli operators $(X_\kappa,Y_\kappa,Z_\kappa)$ with $\kappa=L,R$ 
referring to the left and right box, respectively, using the 
Majorana representation in Eq.~\eqref{PauliOp}. 
 In the present case, it is crucial to
 keep all terms up to third order in the expansion parameters \eqref{condit} when accounting for cotunneling trajectories 
 connecting pairs of QDs, cf.~Fig.~\ref{fig7}. (For the single-box case in Sec.~\ref{sec2a}, 
it is sufficient to go to second order only.)   
The electromagnetic environment then enters the low-energy theory via the three phase differences $\theta_j-\theta_k$ with $j<k$. This fact implies that, in general, we have six different spectral densities ${\cal J}_{jk;j'k'}(\omega)$.   We model these spectral densities by the Ohmic form in Eq.~(\ref{Ohmic}),
with system-bath couplings $\alpha_{jk;j'k'}$. 
For simplicity, we employ an average value $\alpha$ for these couplings below. The bath is then described by a single spectral density ${\cal J}(\omega)$ again.    
 Importantly, the physics is not changed in an essential manner by this approximation. 
In particular, no additional jump operators appear when  allowing for different $\alpha_{jk;j'k'}$. 

\subsubsection{Lindblad equation}

We consider again the weak driving regime with $T\ll \omega_0$. 
Under these conditions, proceeding along similar steps as in Sec.~\ref{sec2b}, one obtains a Lindblad master equation for the density matrix, $\rho(t)$, 
describing both the Majorana sector and the QD degrees of freedom.
In order to arrive at a Lindblad equation for the reduced density matrix, $\rho_{\rm M}(t)$, which refers only to the Majorana sector of both boxes,
we next trace over the QD subsector, see Sec.~\ref{sec2c}.
For the QD steady-state density matrix, $\rho_{\rm d}$, we use the \emph{Ansatz} 
\begin{equation}\label{ansatz2}
\rho_{\rm d}=  {\rm diag}\left(\frac{1-p}{2}, \frac{1-p}{2},p\right), 
\end{equation}
expressed in the basis $\{ |100\rangle,|010\rangle,|001\rangle \}$ with QD occupation states $|n_1,n_2,n_3\rangle$ for $N_{\rm d}=1$.  Note that since we assumed $\epsilon_1=\epsilon_2$, the occupation probabilities of QDs 1 and 2 are equal.
The occupation probability $0<p\ll 1$ refers to the energetically highest QD 3.  Equation \eqref{ansatz2} is consistent with our numerical analysis of the Lindblad equation for
$\rho(t)$, where we again find a factorized density matrix at long times, $\rho(t)\simeq \rho_{\rm M}(t)\otimes \rho_d$.  
We note that the dark space turns out to be independent of the concrete value of $p$.

Going through the corresponding steps in Sec.~\ref{sec2c}, we arrive at a Lindblad equation
for $\rho_{\rm M}(t)$,
\begin{equation} \label{LindbladTwoMBQ}
    \partial_t\rho_{\rm M}(t)=-i[\tilde H_{\rm L},\rho_{\rm M}(t)]+
    \sum_{a=1}^6 \tilde \Gamma_a \mathcal{L}[K_a] \rho_{\rm M}(t).
\end{equation}
The six jump operators are denoted by $K_a$, with the respective dissipative transition rates $\tilde\Gamma_a$. 
With $\lambda_{LR}\equiv t_{LR}/E_C \ll 1$, we obtain
\begin{eqnarray}\nonumber
  K_1^{} &=& K_4^\dagger = ie^{i(\beta_3-\beta_1)}\frac{|\lambda_{1,1R}\lambda_{3,3R}|}{\lambda_{LR}} X_R \\
  &&\qquad - \,e^{i\beta_3} |\lambda_{1,3L}\lambda_{3,3R}|  Z_L Y_R,\nonumber\\ \label{jumpK}
  K_2^{} &=& K_5^\dagger =-ie^{i(\beta_3-\beta_2)}\frac{|\lambda_{2,4R}\lambda_{3,3R}|}{\lambda_{LR}}Z_R\\
  &&\qquad + \, e^{i\beta_3}|\lambda_{2,2L}\lambda_{3,3R}|  X_LY_R,\nonumber\\
  K_3^{} &=& K_6^\dagger =  i\frac{|\lambda_{1,3L}\lambda_{2,2L}|}{\lambda_{LR}}Y_L-
  ie^{i(\beta_2-\beta_1)}
  \frac{|\lambda_{1,1R}\lambda_{2,4R}|}{\lambda_{LR}}  Y_R\nonumber\\ 
    &&\qquad+ \,e^{-i\beta_1} |\lambda_{1,1R}\lambda_{2,2L}| \,X_LZ_R \nonumber \\ \nonumber &&
     \qquad -\,e^{i\beta_2}|\lambda_{1,3L}\lambda_{2,4R}| \,Z_LX_R.
\nonumber
\end{eqnarray}
The coherent  evolution in Eq.~\eqref{LindbladTwoMBQ} is governed by the Hamiltonian 
\begin{equation}\label{h2qdef}
 \tilde H_{\rm L} = 2p\tilde g_0 K_z + \sum_{a=1}^6 \tilde h_a K_a^\dagger K_a,
\end{equation}
with the operator
\begin{equation} \label{kz}
   K_z^{} =\sin\beta_1|\lambda_{1,1R}\lambda_{1,3L}| Z_LZ_R+ 
    \sin\beta_2|\lambda_{2,2L}\lambda_{2,4R}| X_L X_R.
\end{equation}
We here used the energy scale
\begin{equation}\label{tildeg0}
\tilde g_0= \lambda_{LR} g_0 = \frac{t_0^2 t_{LR}}{E_C^2},
\end{equation}
which characterizes the relevant inelastic cotunneling processes in the double-box setup.
The transition rates $\tilde\Gamma_a$ follow in the form
\begin{eqnarray}
\tilde\Gamma_1 &=& \tilde \Gamma_2 = 2p\tilde g_0^2 \,{\rm Re} \int_0^\infty dt e^{i\omega_0 t} e^{J_{\rm env}(t)},\nonumber\\
\tilde\Gamma_3 &=& \tilde \Gamma_6 =(1-p)\tilde g_0^2\, {\rm Re} \int_0^\infty dt e^{J_{\rm env}(t)}   ,\label{trans22}\\
\tilde\Gamma_4 &=&\tilde\Gamma_5= \frac{(1-p)}{2p} e^{-\omega_0/T}\tilde \Gamma_1,\nonumber
\end{eqnarray}
and the Lamb shifts $\tilde h_a$ are given by
\begin{eqnarray}
\tilde h_1 &=& \tilde h_2 = p\tilde g_0^2 \,{\rm Im} \int_0^\infty dt e^{i\omega_0 t} e^{J_{\rm env}(t)},\nonumber\\
\tilde h_3 &=& \tilde h_6 = \frac12 (1-p)\tilde g_0^2 \,{\rm Im} \int_0^\infty dt e^{J_{\rm env}(t)}   ,\label{lamb22}\\
\tilde  h_4 &=&\tilde h_5= \frac{(1-p)}{2p} e^{-\omega_0/T}\tilde h_1.\nonumber
\end{eqnarray} 
For $\omega_0\ll \omega_c$, we can then make further analytical progress.  Explicit 
expressions for $\tilde \Gamma_{1,2}$ and $\tilde h_{1,2}$ follow by comparison with Eq.~\eqref{rate2}.
In addition, we find 
\begin{eqnarray}\nonumber
    \tilde \Gamma_{3,6} &\simeq& (1-p) \frac{\cos(\pi \alpha)\Gamma(\alpha)\Gamma(1-2\alpha)}{2^{1-2\alpha} \Gamma(1-\alpha)}   \left( \frac{\pi T}{\omega_c}\right)^{2\alpha-1} \frac{2g_0^2}{\omega_c} ,\\
    \tilde h_{3,6}&=& -\frac{1}{2}\tan(\pi \alpha) \tilde \Gamma_{3,6}. 
\end{eqnarray}

By following the derivation of the reduced master equation \eqref{LindbladTwoMBQ}, we observe that the operator $K_1$ ($K_2$) comes from unidirectional transitions transferring an electron from the energetically high-lying QD 3 to QD 1 (QD 2) via the double-box setup, collecting all possible cotunneling trajectories allowed by third-order perturbation theory. Likewise, the jump operator $K_4$ ($K_5$) describes the reversed process, with a cotunneling transition from QD 1 (QD 2) to QD 3. 
For $T\ll \omega_0$, the transition rates $\tilde \Gamma_{4,5}$ and Lamb shifts $\tilde h_{4,5}$ are exponentially suppressed, $\propto e^{-\omega_0/T}$, 
against the respective contributions from $K_{1,2}$.
Moreover, the jump operators $K_3$ and $K_6$ in Eq.~\eqref{jumpK} describe cotunneling transitions between  QDs 1 and 2.  Since these QDs are not directly connected by a driven tunnel link and have the same energy, $\epsilon_1=\epsilon_2$, the corresponding rates and Lamb shifts coincide, $\tilde\Gamma_3=\tilde\Gamma_6$ and $\tilde h_3=\tilde h_6$. 
Importantly, for $1/2< \alpha < 1$, these quantities are reduced by a factor $(T/\omega_0)^{2\alpha-1}\ll 1$ against $\tilde\Gamma_{1,2}$ and $\tilde h_{1,2}$, respectively.  
In the remainder of this section, we shall study this parameter regime
where the most important jump operators in Eq.~\eqref{LindbladTwoMBQ} are given by $K_1$ and $K_2$.  Nonetheless, we retain the other jump operators in our numerical analysis as well.

Finally, we note that all terms without the factor $\lambda^{-1}_{LR}\gg 1$ in Eqs.~\eqref{jumpK} and \eqref{kz} stem from third-order processes. While one \emph{a priori} expects that the corresponding dissipative terms in Eq.~\eqref{LindbladTwoMBQ} are suppressed against second-order contributions, by careful tuning of the link transparencies $\lambda_{j,\nu\kappa}$,
they can become of comparable magnitude. 
As a consequence, all relevant cotunneling paths will then have amplitudes corresponding to third-order processes.  This means that for the present two-box setup, the energy scale $g_0=t_0^2/E_C$ appearing in Eq.~\eqref{basiccond} has to be replaced by $\tilde g_0$ in Eq.~\eqref{tildeg0}.
The Lindblad equation \eqref{LindbladTwoMBQ} describing the weak driving limit is 
therefore valid under the conditions
\begin{equation}\label{basiccond2}
\tilde g_0\ll T\ll  \omega_0,\quad A \alt \tilde g_0.
\end{equation}

\subsubsection{Dissipative maps}

Before entering our discussion of 
stabilization protocols for the layout in Fig.~\ref{fig7}, it is convenient to 
introduce the dissipative maps \cite{Barreiro2011}
\begin{equation}\label{bellmap}
\hat E_{1,\pm} = (\mathbb{1}\pm Z_L Z_R) X_R,\quad
\hat E_{2,\pm} = (\mathbb{1}\pm X_L X_R) Z_R.
\end{equation}
These maps can be used to target the four Bell states,
\begin{equation}\label{bellstates}
|\psi_{\pm} \rangle =\frac{1}{\sqrt2}(|00\rangle\pm|11\rangle),
\quad|\phi_{\pm}\rangle=\frac{1}{\sqrt2}(|01\rangle\pm|10\rangle),
\end{equation}
which are eigenstates of both $Z_LZ_R=\pm 1$ and $X_LX_R=\pm 1$.
We observe  that
$\hat E_{1,-}$ maps even-parity onto the respective odd-parity states, 
$\hat E_{1,-}|\psi_{\pm}\rangle=|\phi_{\pm}\rangle$, while  odd-parity states do not evolve in time,
$\hat E_{1,-}|\phi_{\pm}\rangle=0$. 
Under this dissipative map, the system will thus be driven into the degenerate odd-parity subsector spanned by the $|\phi_\pm\rangle$ states.  Similarly, $\hat E_{2,-}$ can drive the system into the antisymmetric subsector spanned by $|\phi_-\rangle$ and $|\psi_-\rangle$.

The key idea in our DD protocols below is to identify state design parameters such that the jump operators effectively 
realize the needed dissipative map(s) in Eq.~\eqref{bellmap}.
Recalling that a dissipative map breaks a number of conserved quantities (and therefore symmetries) in our system, see Refs.~\cite{Albert2014,Albert2016} and 
App.~\ref{appC}, we here employ this insight to either stabilize a dark  space, see Sec.~\ref{sec4b} and Ref.~\cite{ourprl}, or to target  protected and 
maximally entangled two-qubit dark states, see Sec.~\ref{sec4c}.

\subsection{Stabilization of a dark space}\label{sec4b}

In this subsection, we briefly outline how one can stabilize a dark  space in the setup of Fig.~\ref{fig7}, see also Ref.~\cite{ourprl}.  
For convenience, we decouple QD 2 from the system by using the parameter choice
\begin{equation}\label{decouple2}
\lambda_{2,2L}=\lambda_{2,4R}=0,\quad \beta_2=0.
\end{equation} 
We note that this is not the only possible parameter set for constructing a dark space.
As a consequence of Eq.~\eqref{decouple2}, many of the jump operators in Eq.~\eqref{jumpK} vanish identically, $K_2=K_3=K_5=K_6=0$.
The jump operator $K^{}_1=K_4^\dagger$ then yields the dissipative map $\hat E_{1,-}$ in Eq.~\eqref{bellmap} upon choosing 
\begin{equation}
    \beta_1=-\pi, \quad \beta_3 = -\pi/2, \quad |\lambda_{1,1R}|=\lambda_{LR} |\lambda_{1,3L}|.
    \label{spacecond}
\end{equation}
Noting that $\hat E_{1,-}= X_R-i Z_L Y_R$, see Eq.~\eqref{bellmap},  
we indeed arrive at $K_1\propto \hat E_{1,-}$ from Eq.~\eqref{jumpK}.  
In addition, Eq.~\eqref{h2qdef} shows that under the above conditions, $\tilde H_{\rm L}$ only generates terms $\propto Z_L Z_R$ which do not obstruct the dissipative dynamics. 

For $T\ll \omega_0$, we next observe that to exponential accuracy, $K_1$ is the only jump operator contributing to the Lindbladian in Eq.~\eqref{LindbladTwoMBQ}
for the parameters in Eqs.~\eqref{decouple2} and \eqref{spacecond}. 
The DD protocol therefore will stabilize the system in the 
odd-parity ($Z_L Z_R=-1$) Bell state manifold spanned by $\{ |\phi_+\rangle, |\phi_-\rangle \}$.  
We show in Ref.~\cite{ourprl} that this
degenerate manifold has the dark space 
dimension $D=4$, see also App.~\ref{appC}, 
which is equivalent to a degenerate qubit space \cite{Albert2014}. 

It is possible to manipulate dark states within a dark space by following different strategies \cite{ourprl}. 
For instance, one can adiabatically switch on a perturbation that breaks at least one conservation law. 
An alternative possibility is to employ single-electron pumping protocols, in analogy to previous proposals for native Majorana qubits \cite{Plugge2017,Karzig2017}.

\subsection{Stabilizing  Bell states} \label{sec4c}

We next turn to the stabilization of Bell states in the setup of Fig.~\ref{fig7}, where the couplings between QD 2 and the Majorana islands are now assumed finite again.
In that case,  the jump operator $K_2$ in Eq.~\eqref{jumpK} does not vanish anymore.   In the low temperature regime, the corresponding Lindbladian term in Eq.~\eqref{LindbladTwoMBQ}  contributes 
with the same transition rate, $\tilde\Gamma_2=\tilde\Gamma_1$, as for  $K_1$, see Eq.~\eqref{trans22}.
Importantly, $K_2$ breaks additional conservation laws and thereby allows one to engineer stabilization protocols targeting maximally entangled two-qubit states.  We again study the regime $1/2< \alpha<1$, where the jump operators $K_{3,6}$ 
give only subleading contributions.

Let us start with the Bell singlet state $|\phi_-\rangle$ in Eq.~\eqref{bellstates}, where $Z_LZ_R=-1$ and $X_LX_R=-1$. 
By choosing the state design  parameters as
\begin{eqnarray}\label{bellcond}
\beta_1 &=& -\pi,\quad \beta_2 = 0,\quad \beta_3 = -\pi/2,\\
|\lambda_{1,1R}|&=&\lambda_{LR}|\lambda_{1,3L}|,\quad |\lambda_{2,4R}|=\lambda_{LR}|\lambda_{2,2L}|,
\nonumber
\end{eqnarray} 
we observe from Eq.~\eqref{jumpK} that $K_1\propto \hat E_{1,-}$ and $K_2\propto \hat E_{2,-}$ are directly expressed in terms of the corresponding dissipative maps, see Eq.~\eqref{bellmap}.
The Lindbladian will therefore drive the system to the dark state $|\phi_-\rangle$.  The dark space dimension is thus given by $D=1$.

\begin{figure}[t]
\begin{centering}
\includegraphics[width=\columnwidth]{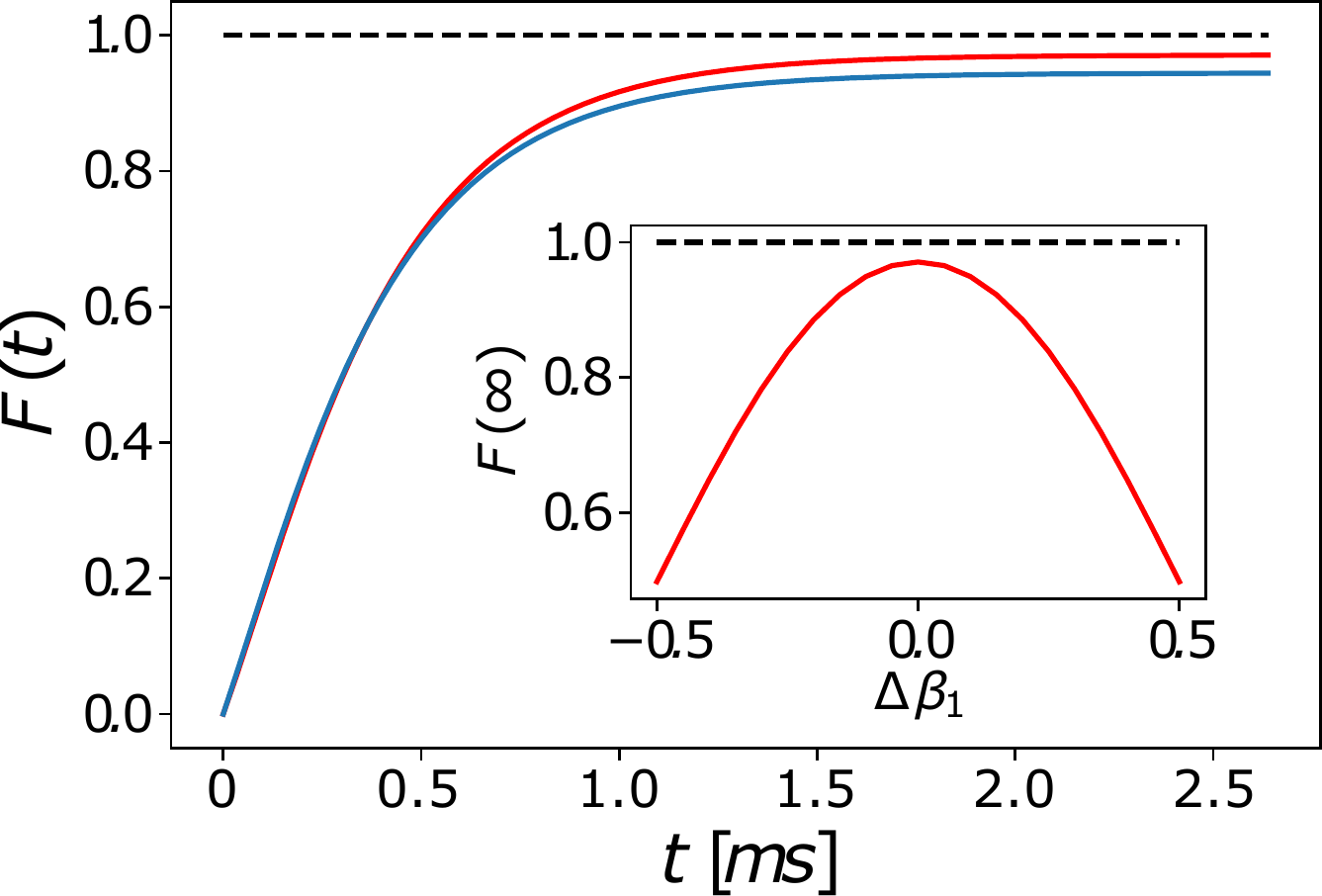}
\end{centering}
\caption{Fidelity for stabilizing the Bell singlet state $|\phi_-\rangle$ in the setup of Fig.~\ref{fig7}.
We show numerical results obtained from Eq.~\eqref{LindbladTwoMBQ} with the parameters 
in Eq.~\eqref{bellcond} and $|\lambda_{1,3L}|=|\lambda_{2,2L}|=|\lambda_{3,3R}|=1$, using the initial state $\rho_{\rm M}(0)=|00\rangle\langle00|$.
Other parameters are $E_C=1$~meV, $\tilde g_0/E_C=10^{-5}$, $T/\tilde g_0=2, \omega_0/\tilde g_0=2\times 10^3, \omega_c/\tilde g_0=10^4, \alpha=0.99$, and $p=0.01$.
Main panel: Time dependence of $F(t)$ for ideal parameters [Eq.~\eqref{bellcond}] (red curve), and for a mismatch of order $10\%$ in all
state design parameters [$|\lambda_{1,1R}|=1.1\lambda_{LR}|\lambda_{1,3L}|, \, |\lambda_{2,4R}|=0.9\lambda_{LR}|\lambda_{2,2L}|, \, \beta_1=-1.1\pi, \, \beta_3=-9\pi/20$] (blue).
Inset: Steady-state fidelity vs deviation $\Delta\beta_1$ from the ideal value, i.e., $\beta_1=-\pi(1+\Delta\beta_1)$,
with otherwise ideal parameters.} 
\label{fig8}
\end{figure}

As is shown in Fig.~\ref{fig8}, the  numerical solution of Eq.~\eqref{LindbladTwoMBQ} confirms this expectation. For the stabilization parameters in Eq.~\eqref{bellcond}, 
the Bell singlet state is reached with nearly perfect fidelity when taking ideal parameter values.
  One can rationalize the almost perfect fidelity by noting that the coherent evolution due to $\tilde H_{\rm L}$, see Eq.~\eqref{h2qdef}, involves only the operators $Z_L Z_R$ and $X_L X_R$. As a consequence, the dynamics induced by the dissipative maps $K_{1,2}\propto \hat E_{1/2,-}$  will not be disturbed. 
Note that the parameters in Fig.~\ref{fig8} were chosen such that $\tilde\Gamma_1\gg\tilde\Gamma_3$ while staying in the regime specified in Eq.~\eqref{basiccond2}.  Indeed,
the observed small deviations from the ideal value $F=1$, see Fig.~\ref{fig8}, can be traced back to the jump operators $K_3$ and $K_6$, which give nominally subleading but practically important contributions to the Lindblad equation.
 
 Figure \ref{fig8} shows that the stabilization protocol is rather robust against 
deviations of state design parameters from their ideal values in Eq.~\eqref{bellcond}, see Sec.~\ref{sec3}. 
 Following the approach in App.~\ref{appB}, we find that the dissipative gap for stabilizing $|\phi_-\rangle$ is given by
\begin{equation}\label{bellgap}
    \Delta_{\rm Bell}=|2\lambda_{3,3R}|^2\left(|\lambda_{1,3L}|^2+|\lambda_{2,2L}|^2\right)\sum_{a=1,2,4,5}\tilde\Gamma_a.
\end{equation}
Due to the importance of third-order inelastic cotunneling processes, this dissipative gap is several orders of magnitude below the 
corresponding gaps in the single-box case, cf.~Sec.~\ref{sec3}. 
For the parameters in Fig.~\ref{fig8}, we obtain the time scale $\Delta_{\rm Bell}^{-1}\approx 0.3$~ms.

The other Bell states in Eq.~\eqref{bellstates} can be targeted by changing the  phases $\beta_j$ in Eq.~\eqref{bellcond}.  The jump operators $K_1$ and $K_2$ will then directly implement the desired dissipative maps, with the dissipative gap still given by Eq.~\eqref{bellgap}.  For stabilization of the Bell state $|\psi_+\rangle$ ($|\psi_-\rangle$), one has to 
put $\beta_1=0, \, \beta_2=\pi \, (\beta_2=0)$, and $\beta_3=\pi/2$.  Similarly,  $|\phi_+\rangle$ is stabilized for $\beta_1=-\pi, \beta_2 = \pi,$ and $\beta_3= -\pi/2$. We thus always have $\beta_3-\beta_1=\pi/2$, and the remaining two phases select the targeted Bell state.
In particular, $\beta_1$ selects the parity of the target state while $\beta_2$ determines the symmetric vs antisymmetric state.

\section{Summary and prospects}\label{secConc}

In this paper, we have described DD protocols in Majorana-based layouts for stabilizing as well as manipulating dark states and dark spaces. 
For devices with one or two Majorana boxes coupled to driven QDs and subject to electromagnetic noise, we have shown that 
in a wide parameter regime the dynamics in the Majorana sector is accurately described by Lindblad master equations.

The underlying topological nature of the Majorana states significantly boosts the power of DD schemes in several directions. 
  First, the role of uncontrolled environmental noise sources should be suppressed compared to topologically trivial realizations, which is a key advantage for high-dimensional dark space constructions.  
  Second, the fact that Pauli operators describing native Majorana qubits correspond to products of Majorana operators (pertaining to spatially separated MBSs), see Eq.~\eqref{PauliOp}, allows for unique addressability options.
  Only through this feature, which is rooted in topology, it is possible to design the special unidirectional cotunneling paths which directly
   implement the jump operators appearing in the Lindblad equation.  In the simplest single-box case, see Fig.~\ref{fig1}, the basic pumping-cotunneling cycle involves 
   (i) pumping the dot electron from QD 1 to the high-lying QD 2 by means of a weak driving field, and (ii) the back transfer of the electron from QD 2 to QD 1 by cotunneling through the 
   box.  In general, competing transfer mechanisms may also contribute to both steps, and the parameter regime has to be carefully adjusted to 
   minimize their impact.  Taking step (ii) as example, the drive Hamiltonian in Eq.~\eqref{Hdriv}, possibly together with photon emission processes, may provide such
   a competing rate.  By choosing both a sufficiently small drive amplitude, $A<g_0$, and a very small direct tunnel coupling $t_{12}$ between both QDs,
   these competing rates can be systematically suppressed against the cotunneling rates through the box.
We also note that in most cases of interest, the Lindbladian dissipator alone is responsible for driving the system into the desired dark state or dark space, i.e., the Hamiltonian appearing in coherent part of the Lindblad equation does not obstruct the dissipative dynamics.

For a single-box architecture, we have shown how to stabilize arbitary pure dark states, i.e., states that are fault tolerant and stable on arbitrary time scales.  For multiple-box devices, one can also stabilize dark spaces, i.e., manifolds of degenerate dark states, as well as protected two-qubit Bell states.  In our accompanying short paper \cite{ourprl}, we show that a two-box device allows one to implement a dark Majorana qubit, which in turn could serve as basic ingredient for dark space quantum computation schemes.   
 Our stabilization and manipulation protocols can be implemented with available hardware elements once a working Majorana platform becomes available.

The above concepts and ideas raise many interesting perspectives for future research.   First, we expect that one can devise robust 
Majorana braiding protocols \cite{Alicea2012,Leijnse2012,Beenakker2013} that are stabilized by working within a dark space manifold.
Second, for chains of many boxes, our DD stabilization protocols may allow for interesting quantum simulation
applications, e.g., a realization of the topologically nontrivial ground state of spin ladders \cite{Ebisu2019} or of the
Affleck-Kennedy-Lieb-Tasaki (AKLT) spin chain \cite{Kraus2008,Affleck1987}. 
For clarifying the feasibility of such ideas, one needs to analyze the spectrum of the Lindbladian for DD multiple-box networks.   We leave this endeavor to future work.

\begin{acknowledgments} 
We thank A. Altland, S. Diehl, and K. Snizhko for discussions.
This project has been funded by the Deutsche Forschungsgemeinschaft (DFG,
German Research Foundation) under Grant No.~ 277101999, TRR 183 (project
C01), under Germany's Excellence Strategy - Cluster of Excellence Matter
and Light for Quantum Computing (ML4Q) EXC 2004/1 - 390534769, and under Grant No.~EG 96/13-1. In addition, we acknowledge funding  
 by the Israel Science Foundation.  
\end{acknowledgments}

\appendix

\section{On the strong driving limit}\label{appA}

We here briefly discuss the strong driving limit for the single-box device in Fig.~\ref{fig1}, with
total QD occupancy $N_{\rm d}=1$ and under resonant driving conditions, $\omega_0=\epsilon_2-\epsilon_1$.   We consider the regime
\begin{equation}
  g_0 \ll T<A\ll  \omega_0,
\end{equation}
with otherwise identical conditions as in Sec.~\ref{sec2}. 
After imposing the RWA, the steady-state density matrix of the QDs is given by Eq.~\eqref{ssform} with $p=1/2$ and $p_\perp=0$.

Starting from the effective Hamiltonian $H_{\rm eff}(t)$ in Eq.~\eqref{Hfull},
we then arrive at a Lindblad equation for the density matrix $\rho(t)$ describing the combined system of MBSs and QDs,
\begin{equation}\label{AAA}
\partial_t \rho(t) = -i [ H_{\rm L}, \rho(t) ] +
2 g_0^2 \sum_{a = 1}^3 \sum_{s = \pm} {\rm Re}\,\Lambda_{a,s} \, {\cal L} [ J_{a,s} ] \rho(t),
\end{equation}
with the effective Hamiltonian
\begin{equation}
H_{\rm L} = A\tau_x + g_0^2  \sum_{a = 1}^3 \sum_{s = \pm} {\rm Im}\,\Lambda_{a,s} \, J_{a,s}^\dagger J_{a,s}^{}.
\end{equation}
We here encounter \emph{six} jump operators ($s=\pm$),
\begin{equation}\label{rwa:Jaq}
J_{1,s} = \tilde J_s \tau_x,\quad J_{2,s} = J_{3,-s}^\dagger =
\tilde J_s (\tau_z + i \tau_y )/2, 
\end{equation}
with the operators $\tilde J_\pm$ in Eq.~\eqref{jumpN}.
 The dissipative transition rates as well as the Lamb shifts follow from  $\Lambda_{1,\pm}\equiv \Lambda_\pm$, see Eq.~\eqref{lambdadef}, and 
\begin{equation}\label{rwa:Lambdaq}
\Lambda_{2/3,s} = \int_0^\infty dt \, e^{i s \omega_0 t \pm i A t} e^{J_{\rm env}(t)},
\end{equation}
with the bath correlation function \eqref{BathCorr}.
Comparing to the weakly driven case in Sec.~\ref{sec2b}, the strong driving field $A$ splits the two jump operators $J_s$ in Sec.~\ref{sec2b} into the six jump operators in Eq.~\eqref{rwa:Jaq}.

Tracing over the QD degrees of freedom, we arrive at a Lindblad equation for the density matrix $\rho_{\rm M}(t)$, cf.~Sec.~\ref{sec2c},
\begin{equation}\label{rwa:LBM}
\partial_t \rho_{\rm M}(t)  = - i [ \tilde H_{\rm L}, \rho_{\rm M}(t) ] +
\sum_{s = \pm} \tilde\Gamma_s \, {\cal L }[ \tilde J_s ] \rho_{\rm M}(t),
\end{equation}
with
$\tilde H_{\rm L} = {\rm Tr}_{\rm d} \left\{ \rho_{\rm d} H_{\rm L} \right\}$. 
In this expression, $\rho_{\rm d}$ follows from Eq.~\eqref{ssform} with $p\to 1/2$ and $p_\perp \to 0$.  Only the two jump operators $\tilde J_\pm$ appear in the reduced Lindblad equation \eqref{rwa:LBM}, with the dissipative transition rates 
\begin{equation}\label{rwa:Gammaq}
\tilde\Gamma_s =2 g_0^2 \, {\rm Re}\left[\Lambda_{1,s} +\frac12
\left( \Lambda_{2,s} + \Lambda_{3,s} \right) \right].
\end{equation}
Finally, we note that for $T>A\gg g_0$, the Lindblad equation \eqref{LindbladMBQ} holds with $p \to 1/2$.

\section{On the dissipative gap}\label{appB}

An elegant way to study the spectrum of a general Lindbladian uses the so-called Choi isomorphism in order to map the $N\times N$ system density matrix, $\rho(t)$, 
to an $N^2\times 1$ vector, $|\rho(t)\rangle$, and the Liouvillian, $\mathcal{\hat L}$, to an $N^2\times N^2$ superoperator ${\bm L}$ \cite{Albert2014}. 
We here include the Hamiltonian part in $\mathcal{\hat L}$.

Let us consider a general Lindblad master equation, cf.~Eq.~\eqref{Dissipator},
\begin{equation}\label{genB1}
\partial_t\rho(t)=\mathcal{\hat L}\rho(t)=-i[ H,\rho(t)]+\sum_{a}\Gamma_{a}{\cal L} [ J_{a}]  \rho(t),
\end{equation}
with jump operators $J_a$ and the corresponding transition rates $\Gamma_a$. Using the isomorphism, 
we have the correspondence $J\rho J^\dagger \leftrightarrow ( J\otimes J^*)|\rho\rangle$,
and Eq.~\eqref{genB1} takes the equivalent form $\partial_t|\rho(t)\rangle={\bm L}|\rho(t)\rangle$ with
\begin{eqnarray}\label{genB2}
{\bm L}&=&-i\left(H\otimes \mathbb{1}- \mathbb{1}\otimes H^\ast\right)+ \sum_{a}\frac{\Gamma_{a}}{2} \times \\
\nonumber &\times&
\left(2 J_{a}^{}\otimes J^{\ast}_{a}-\mathbb{1}\otimes \bigl(J_a^{\dagger} J_{a}^{}\bigr)^\ast  
-J_{a}^{\dagger} J_{a}\otimes\mathbb{1}\right).
\end{eqnarray}
In this language, the steady state, $\rho_{\rm ss}$, follows as right eigenvector of ${\bm L}$ with eigenvalue zero, 
\begin{equation}
{\bm L}\left|\rho_{\rm ss}\right\rangle=0.
\label{Criterion}
\end{equation}
Equation \eqref{Criterion} allows one to systematically search for stabilization conditions targeting a desired dark state. Moreover, the spectrum of the Lindbladian 
coincides with the eigenvalues of the superoperator $\bm L$. In particular, the number of zero eigenvalues defines the
dark space dimension, $D$, and the dissipative gap equals the real part of the smallest non-zero 
eigenvalue \cite{Albert2014}.

\section{On conserved quantities}\label{appC}

For an open quantum system described by a Lindbladian as in Eq.~\eqref{genB1}, where we assume that $\mathcal{\hat L}$ has no purely imaginary eigenvalues, 
it is known that all conserved quantities are linked to the basis states spanning the dark space \cite{Albert2014}. For a Lindbladian with 
$D$ conserved quantities $C_{\mu=1,\ldots,D}$,  we have the commutation relations  
\begin{equation}
    [H, C_\mu]=[J_a, C_\mu] = 0.
\end{equation}
Using an orthonormal basis, $\lbrace M_\mu\rbrace_{\mu=1}^D$, to span the resulting $D$-dimensional dark space, the 
steady state can be written as 
\begin{equation}
    \rho_{\rm ss} = \underset{t\to \infty}{\rm lim} e^{\mathcal{\hat L}t} \rho(0)=\sum_{\mu=1}^D c_\mu M_{\mu},
    \label{generaldarkspace}
\end{equation}
where $\rho(0)$ is the initial density matrix and the $c_\mu={\rm tr}[C_\mu^{\dagger} \rho(0)]$ 
are weights determining in which of the degenerate steady states the system ends up.

As first illustration, let us consider the stabilization of the dark state $|0\rangle$ for a single-box device, cf.~Sec.~\ref{sec3a} and Eq.~\eqref{sigmazcond}.
The jump operators are then given by $\tilde J_\pm\propto \sigma_\pm$.
The only operator commuting with both $\tilde J_+$ and $\tilde J_-$ is the identity, $C_\mu = \mathbb{1}$, and thus the dark space dimension
is $D=1$.   For this example, we also have $H=\tilde H_{\rm L}\propto Z$, see Eq.~\eqref{mod22}. We conclude that $M_1=|0\rangle\langle 0|$ spans the corresponding space.

As second example, we discuss the dark space stabilization for a two-box device in Sec.~\ref{sec4b}. 
Using the Lindblad equation \eqref{LindbladTwoMBQ} and assuming that QD $2$ remains decoupled from the system, see Eq.~\eqref{decouple2},   
the four conserved quantities  $C_{\mu}$ listed in Ref.~\cite{ourprl} are readily identified. 
Given these quantities,  a basis spanning the dark space can be constructed from Eq.~\eqref{generaldarkspace}. 
One may view the basis elements, $M_{\mu}$, as linearly independent `vectors' with the orthogonality relation
${\rm tr}( M_{\mu}^{\dagger}M_{\nu}^{})=\delta_{\mu\nu}$. 
The existence of four conserved operators $C_{\mu}$ now implies that we have four basis vectors spanning the dark space, 
see Ref.~\cite{ourprl} for explicit expressions. Since the dark space dimension $D$ coincides with the number of linearly independent basis vectors, 
we have $D=4$ for the case studied in Sec.~\ref{sec4b} and Ref.~\cite{ourprl}.  
Since the $C_{\mu}$ and $M_{\mu}$ specified in Ref.~\cite{ourprl} form the Lie algebra u$(2)$ \cite{Albert2014},
this dark space is equivalent to a degenerate qubit space.

\end{document}